\documentstyle[12pt]{article}
\newcommand{\beq}{\begin{equation}}
\newcommand{\eeq}{\end{equation}}
\newcommand{\beqa}{\begin{eqnarray}}
\newcommand{\eeqa}{\end{eqnarray}}
\newcommand{\beqar}{\begin{eqnarray*}}
\newcommand{\eeqar}{\end{eqnarray*}}
\newcommand{\htx}{s} 
\newcommand{\al}{\alpha}

\renewcommand{\l}{\lambda}

\newcommand{\ssc}{\scriptscriptstyle}
\newcommand{\eg}{{\it e.g.,}\ }
\newcommand{\ie}{{\it i.e.,}\ }

\newcommand{\hR}{\hat{R}}

\newcommand{\norm}[1]{\raise.3ex\hbox{:}#1\raise.3ex\hbox{:}}

\newcommand{\labell}[1]{\label{#1}} 
\newcommand{\labels}[1]{\label{#1}} 
\newcommand{\reef}[1]{(\ref{#1})}
\newcommand{\Tr}{{\rm Tr}}
\newcommand{\STr}{{\rm STr}}
\newcommand{\NN}{{\rm N}} 

\newcommand\hi{{\rm i}}
\newcommand\hPhi{{\hat\Phi}}
\newcommand\prt{\partial}
\newcommand\ls{\ell_s}

\newcommand\tB{{\widetilde B}}

\newcommand\cL{{\cal L}}
\newcommand\vareps{\varepsilon}
\newcommand\sig{\sigma}
\begin{document}

\thispagestyle{empty}
\rightline{\small hep-th/9910053 \hfill McGill/99-27}
\rightline{\small \hfill NSF-ITP/99-113}
\vspace*{2cm}

\begin{center}
{\bf \LARGE Dielectric-Branes}
\vspace*{1cm}

Robert C. Myers\footnote{E-mail: rcm@hep.physics.mcgill.ca}
\vspace*{0.2cm}

{\it Institute for Theoretical Physics, University of California}\\
{\it Santa Barbara, CA 93106 USA}\\[.6em]
{\it Department of Physics, McGill University}\\
{\it Montr\'eal, QC, H3A 2T8, Canada}\footnote{Permanent Address}\\

\vspace{2cm} ABSTRACT
\end{center}
We extend the usual world-volume action for a D$p$-brane
to the case of N coincident D$p$-branes where the world-volume
theory involves a U(N) gauge theory. The guiding principle in our
construction is that the action should be consistent with
the familiar rules of T-duality. The resulting action involves
a variety of potential terms, {\it i.e.,} nonderivative interactions,
for the nonabelian scalar fields. This action also shows that
D$p$-branes naturally couple to RR potentials of all form degrees,
including both larger and smaller than $p$+1. We consider the
dynamics resulting from this action for D$p$-branes moving in
nontrivial background fields, and 
illustrate how the D$p$-branes are ``polarized'' by external fields.
In a simple example, we show that a system of D0-branes in an
external RR four-form field expands into a noncommutative two-sphere,
which is interpreted as the formation of a spherical D2-D0 bound state.

\vfill \setcounter{page}{0} \setcounter{footnote}{0}
\newpage

\section{Introduction}

Our understanding of string theory has been transformed since
1995 \cite{parad}. One important aspect in this transformation
was the appreciation of the important role of extended objects
beyond strings in these theories. Of particular interest
for the type II (and I) superstring theories are the Dirichlet-branes
(D-branes) which can be regarded as stringy solitons
which carry Ramond-Ramond (RR) charges \cite{Polchin}
--- see also \cite{Polchin2}. 

These objects arise in considering T-duality
transformations of a ten-dimensional type I superstring theory
\cite{horava,leigh,green}. So
as originally conceived, a D$p$-brane is a
($p+1$)-dimensional extended surface in spacetime which supports
the endpoints of open strings.
The massless modes of this open string theory form a supersymmetric
U(1) gauge theory with a vector $A_a$, $9-p$ real scalars $\Phi^i$ and 
their superpartner fermions. 
At leading order, the low-energy action corresponds to the
 dimensional reduction of that for ten-dimensional
U(1) super-Yang-Mills theory. As usual in string theory, there are
higher order $\alpha'=\ls^2$ corrections, where $\ls$ is the string length
scale. Following the work of ref.~\cite{callan}, Leigh showed that the action
incorporating these corrections to all orders in the field strength takes the
Born-Infeld form \cite{bin} 
\beq
S_{BI}=-T_p \int d^{p+1}\sig\ \left(e^{-\phi}\sqrt{-det(P[G+B]_{ab}+
2\pi\ls^2\,F_{ab})}\right)
\labell{biact}
\eeq
where $T_p$ is the brane tension. The higher order corrections in
this action can be trusted as long as derivatives
of the gauge field strength (and second derivatives of the scalars)
are small on the string scale $\ls$. Leigh's action \reef{biact}
also reveals that the D-branes are dynamical objects,
whose transverse displacements\footnote{Here and in the
following equation, we have chosen static gauge --- see section 2.} 
are described by $\Phi^i$, \ie $\Delta X^i(\sig)=2\pi\ls^2 \Phi^i$.
This dynamics is implicit in the pull-back of the
bulk spacetime tensors to the D-brane world-volume, which
is denoted by the symbol $P[\ldots]$, \eg 
\beq
P[G]_{ab}=G_{ab}+4\pi\ls^2\,G_{i(a}\,\prt_{b)}\Phi^i+
4\pi^2\ls^4\,G_{ij}\prt_a\Phi^i\prt_b\Phi^j\ .
\labell{pull}
\eeq
Eq.~\reef{biact} incorporates
the couplings of the world-volume vector and scalars to the
massless Neveu-Schwarz fields of the bulk closed string theory,
\ie the metric, dilaton and Kalb-Ramond field. 
Since a D-brane also carries an RR charge \cite{Polchin}, there
must be couplings to the massless RR states of the closed string.
These interactions are incorporated in
a Chern-Simons action of the form\cite{mike,cs}
\beq
S_{CS}=\mu_p\int P\left[ \sum C^{(n)}\,e^B\right]e^{2\pi\ls^2\,F}
\labell{csact}
\eeq
where $C^{(n)}$ denote the ($n$+1)-form RR potentials
and $\mu_p$ is the RR charge of the brane. Thus a D$p$-brane is naturally
charged under the ($p$+1)-form RR potential. However in the presence
of background Kalb-Ramond fields or world-volume gauge fields, it may
also carry a charge of the RR potentials with a lower form degree,
as allowed with the couplings induced by the exponential factor\cite{mike}.
Such configurations have an interpretation in terms of bound states
of D-branes of different dimensions\cite{bound}. More exotic gauge
field configurations also have interesting interpretations in terms
of intersecting branes\cite{bion}.
Note that in general the background spacetime fields are functions of the
transverse coordinates, and so in the above actions
\reef{biact} and \reef{csact}, they become functionals of the scalars
$\Phi^i$. Hence, even in the leading low energy approximation,
these scalars would be governed by a non-linear sigma model action.

One of the most remarkable properties of D-branes is that the
U(1) gauge symmetry of an individual D-brane is enhanced to a
nonabelian U(N) symmetry for N coincident D-branes \cite{bound}.
As N parallel D-branes approach each other, the ground state modes of
strings stretching between the D-branes become massless. These
extra massless states carry the appropriate charges to then fill
out U(N) representations and the U(1)$^{\rm N}$ of the individual D-branes
is enhanced to U(N). Hence $A_a$ becomes a nonabelian gauge field
and the scalars $\Phi^i$ become scalars in the adjoint representation
of U(N). Understanding how to accommodate this simple yet remarkable
modification in the world-volume actions for general backgrounds
has only received limited attention\cite{moremike,extra}.

The naive extension of the Chern-Simons action \reef{csact}
is apparently straightforward. One would include an additional
trace over gauge indices of the nonabelian field strength which 
now appears in the exponential factor. However,
Douglas\cite{moremike} has proposed
that the background fields should be functionals of the non-abelian scalars
(rather than, \eg only the U(1) or center-of-mass component of $\Phi^i$).
Further it was pointed out \cite{hull} that the pull-backs of the
bulk spacetime tensors should be defined in terms of covariant derivatives
of the full nonabelian scalar fields. That is,
$\prt_a\Phi^i\rightarrow D_a\Phi^i$ in eq.~\reef{pull}. Both of these
suggestions were confirmed to leading order by examining string
scattering amplitudes \cite{scatt,scatt2}. Hence it would seem that
extending the Chern-Simons action to accomodate the nonabelian
world-volume theory involves simply introducing a gauge trace
which encompasses all of the fields (and the pull-back) appearing
in eq.~\reef{csact}. However, in the following section, we will argue
that the resulting action is incomplete as it is not consistent
with T-duality. Extending the action to one consistent with T-duality
will involve introducing extra terms involving commutators of
the nonabelian scalars. This construction reveals that D$p$-branes also have
natural couplings to the RR potentials with form degree larger
than $p$+1.
 
We also use T-duality to build a consistent nonabelian extension of
the Born-Infeld action \reef{biact}. Our starting point is
the D9-brane action which contains no scalars, and we simply
apply a T-duality transformation on $9-p$ directions to produce
that for a D$p$-brane. The latter also generates new terms involving
commutators of the nonabelian scalars. In incorporating the
U(N) gauge symmetry here, we also include
a single gauge trace\footnote{Of course, there is some ambiguity
as to the precise ordering of quantities inside this trace \cite{yet,yet2},
and we are only partially able to resolve this question below.}
encompassing not only the explicit gauge fields
and commutators, but also on the implicit nonabelian scalars in the
functional dependence and pull-backs of the background fields
\cite{moremike,hull}.
The appearance of the latter can again be confirmed for the Born-Infeld
part of the action by examining string scattering amplitudes
\cite{scatt,scatt2}.

An outline of the paper is as follows: We begin in section
\ref{prelim} by establishing our conventions and also recalling the T-duality
transformations of both the spacetime and world-volume fields.
We argue here that introducing a single gauge trace and substituting
nonabelian fields everywhere in eqs.~\reef{biact} and \reef{csact}
yields an incomplete action.
In section \ref{act}, we apply T-duality to construct a consistent
nonabelian extension of the Born-Infeld action. In section \ref{coup},
we present a similar construction for the Chern-Simons part
of the action. In section \ref{cheque}, we discuss some consistency
checks for the resulting actions. In particular, we compare the
linear interactions of closed string fields to the nonabelian D0-brane action
with the interactions derived for this case from matrix theory \cite{wati1}.
In section \ref{diel}, we begin an investigation of the
dynamics resulting from our action for D$p$-branes moving in a
nontrivial background fields, and 
discuss how the D$p$-branes can be ``polarized'' by external fields.
In a simple example, we show that a system of D0-branes in an
external RR four-form field expands into a noncommutative two-sphere,
which is interpreted as the formation of a spherical D2-D0 bound state.
We also compare these results with those calculated in the
dual picture of a D2-brane carrying a nonvanishing U(1) field strength.
The final section provides some discussion of our results.

\section{Preliminaries} \labels{prelim}

As we are interested in the dynamics of D-branes in nontrivial
background fields, we begin by establishing our conventions for
the massless bosonic fields in the type II superstring theories.
The bosonic part of the low-energy action for type IIa string theory
in ten dimensions may be written
as (see \eg \cite{Ortin})\footnote{Further,
the RR 9-form potential may be introduced to
extend this (massless) IIa supergravity to the massive Romans
supergravity\cite{larry} --- see also \cite{moremass}.}
\beqa
I_{IIa} &=&
{1\over2\kappa^2}\int\!d^{10}\!x \sqrt{-G}
\left\{ e^{-2 \phi}\left[R + 4
( \nabla \phi)^2 -{1 \over 12}H^2\right] -
{1 \over 4} (F^{(2)})^2
\right.\nonumber\\
&&\qquad\qquad\qquad\qquad\left.- {1 \over
48}(F^{(4)})^2\right\} -{1\over4\kappa^2}\int B
dC^{(3)}dC^{(3)}
\labell{actionA}
\eeqa
where $G_{\mu\nu}$ is the string-frame metric, $\phi$ is the dilaton,
$H=dB$ is the field strength of the Neveu-Schwarz two-form, 
while the Ramond-Ramond field strengths are
$F^{(2)}=dC^{(1)}$ and $F^{(4)}=dC^{(3)}+H\,C^{(1)}$.

For the type IIb theory, we write the low-energy action as
\beqa
I_{IIb} &=& {1\over2\kappa^2}\int\!d^{10}\!x \sqrt{-G}\left\{
e^{-2 \phi}\left[R + 4( \nabla \phi)^2
-{1 \over 12} H^2\right] 
- {1 \over 12} (F^{(3)} + C^{(0)} H)^2 \right.
\nonumber\\
&&\left.\qquad\qquad
-{1\over2}(\prt C^{(0)})^2-{1\over 480}(F^{(5)})^2\right\}
+{1\over4\kappa^2}\int \left(C^{(4)}+{1\over2}B\,C^{(2)}\right)\,F^{(3)}\,H
\labell{actionB}
\eeqa
where the notation for the Neveu-Schwarz fields is the same as
above, while $F^{(3)}=dC^{(2)}$ and $F^{(5)}=dC^{(4)}+H\,C^{(2)}$
are RR field strengths, and $C^{(0)}$ is the RR scalar.
We are adopting the convention here that the the self duality constraint
$F^{(5)}=\,^*\!F^{(5)}$ is imposed by hand at the level of the equations
of motion\cite{boon}. 

Given these actions, one may construct low energy background 
field solutions corresponding to various D-brane configurations
--- see, for example, refs.~\cite{solute}. Note, however, that
our conventions for the RR fields may differ slightly
from the common choices in the supergravity literature.\footnote{For
example, translating to the fields in \cite{useful} requires:
\vskip -.4cm
\[
(C^{(0)},C^{(1)},C^{(2)},C^{(3)},C^{(4)}) =
(\chi=A^{(0)},-A^{(1)},A^{(2)},A^{(3)},A^{(4)}-{1\over2}B\,A^{(2)})\ .
\]}
The present choice coincides with the RR potentials appearing
in the Chern-Simons action \reef{csact}. This may be confirmed
by verifying that the gauge invariances of the low-energy action
are identical with those of world-volume D-brane action. Further
in the action, there is an overall factor of $(2\kappa^2)^{-1}$
which in particular multiplies the terms involving the RR fields.
While this convention is standard in the literature on solutions
of the supergravity equations
(or some discussions of S-duality, \eg refs.~\cite{john,arkady}
or matrix theory, \eg ref.~\cite{wati1}),
it is an unusual normalization
of the RR fields compared to the standard discussions of D-brane
physics --- compare to ref.~\cite{Polchin2}. 
With regards to this point, our convention will be that the dilaton 
$\phi$ vanishes asymptotically, and then the ten-dimensional Newton's constant
is given by $2\kappa^2=16\pi G_N=(2\pi)^7\ls^8g^2$ where $g$
is the (asymptotic) closed string coupling.
Finally, these actions are presented in terms
of the string-frame metric, and as usual converting to the
Einstein-frame metric is accomplished with
\beq
g_{\mu\nu}=e^{-\phi/2}\,G_{\mu\nu}\ .
\labell{metrics}
\eeq

Now we wish to recall the T-duality transformations of these
supergravity fields. T-duality acts on the
Neveu-Schwarz fields as \cite{bush}:
\beqa
\tilde{G}_{ yy} &=& {1\over G_{yy}}
\qquad\qquad\qquad\qquad
\qquad\qquad\qquad\qquad
e^{2 \tilde{\phi}} =\, { e^{2 \phi} \over G_{yy}}
\nonumber\\
\tilde{G}_{ \mu \nu} &=& G_{\mu \nu}
- { G_{\mu y} G_{\nu y}
- B_{\mu y} B_{\nu y}
\over G_{ yy}}
\qquad\qquad\qquad\quad
\tilde{G}_{\mu y} ={ B_{\mu y}
\over G_{yy}}
\labell{NSrule}\\
\tB_{ \mu \nu}&=&B_{ \mu \nu}
-{B_{\mu y} G_{\nu y}-G_{\mu y}
B_{\nu y}\over G_{yy}}
\qquad\qquad\qquad\quad
\tB_{\mu y} ={ G_{\mu y}
\over G_{ yy}}
\nonumber
\eeqa
Here $y$ denotes the Killing coordinate with respect to which
the T-dualization is applied, while $\mu,\nu$ denote any
coordinate directions other than $y$. If $y$ is identified on a circle
of radius $R$, \ie $y\sim y+2\pi R$, then after T-duality
the radius becomes $\tilde{R}=\alpha'/R=\ls^2/R$. The string coupling
is also shifted as $\tilde{g}=g\ls/R$. 

In section \ref{act}, it will be useful to implement T-duality
on several Killing directions in a single transformation.
Such a transformation is most easily implemented by first
defining the ten-by-ten matrix\cite{revue}
\beq
E_{\mu\nu}=G_{\mu\nu}+B_{\mu\nu}\ .
\labell{eee}
\eeq
Then acting with T-duality on a set of directions denoted
with $i,j=p+1,\ldots,9$, the fields transform as: 
\beq
\tilde{E}_{ab}=E_{ab}-E_{ai}E^{ij}E_{jb}
\ ,\qquad
\tilde{E}_{aj}=E_{ak}E^{kj}
\ ,\qquad
\tilde{E}_{ij}=E^{ij}
\labell{transformed}
\eeq
where $a,b=0,1,\ldots,p$ denote the remaining coordinate directions.
Here $E^{ij}$ denotes the inverse of $E_{ij}$, \ie
$E^{ik}E_{kj}=\delta^i{}_j$. 
One also has the dilaton transformation
\beq
e^{2\tilde\phi}=e^{2\phi}\,det(E^{ij})\ .
\labell{dilaton}
\eeq
Of course, similar statements as above apply with regards
to transforming the periodicities of the $x^i$ directions and the
string coupling constant $g$.

Recall that T-duality transforms the type IIa theory into the
type IIb theory and vice versa, through its action on the world-sheet
spinors\cite{huet,leigh}. This aspect of T-duality is then apparent
in the transformations of the RR fields. The odd-form potentials of the
IIa theory are traded for even-form potentials in the IIb theory
and vice versa. Using the conventions adopted above,
the transformation rules for the RR potentials are \cite{meesort}:
\beqa
\tilde{C}^{(n)}_{\mu\cdots\nu\al y}&=&
C^{(n-1)}_{\mu\cdots\nu\al}-(n-1)
{C^{(n-1)}_{[\mu\cdots\nu| y}G_{|\al]y}\over G_{yy}}
\labell{RRrule}\\
\tilde{C}^{(n)}_{\mu\cdots\nu\alpha\beta}&=&
C^{(n+1)}_{\mu\cdots\nu\alpha\beta y}
+nC^{(n-1)}_{[\mu\cdots\nu\alpha}B_{\beta]y}
+n(n-1){C^{(n-1)}_{[\mu\cdots\nu|y}B_{|\alpha|y}G_{|\beta]y}\over
G_{yy}}
\nonumber
\eeqa
where in general one would have $C^{(n)}$ with $n=0,2,4,6,8,10$ for the
type IIb theory, and $n=1,3,5,7,9$ for type IIa theory --- $n=9$ and 10 
being exceptional cases\cite{Polchin}. This range of $n$ is consistent
with the implicit summation over RR potentials in the world-volume
action \reef{csact}. However, recall that these are not all independent,
rather they appear in dual pairs (for $n\le8$) where at linear order
$dC^{n}=(-)^{n(n-1)/2}dC^{(8-n)}$. Of course, this is why the
low-energy supergravity theories \reef{actionA} and \reef{actionB}
are written in terms of RR potentials with only $n\le4$. 
The T-duality transformations \reef{RRrule} are
consistent with those derived from the supergravity
actions \cite{Ortin}.

We now turn to consider the  D$p$-branes.
To begin, we remark that throughout we are employing static gauge. That
is first we employ spacetime diffeomorphisms to define the fiducial
world-volume as $x^i=0$ with $i=p+1,\ldots,9$, and then with
world-volume diffeomorphisms, we match the internal coordinates with
the remaining spacetime coordinates on that surface,
$\sig^a=x^a$ with $a=0,1,\ldots,p$. With this choice and the identification
$x^i=2\pi\ls^2\Phi^i$, the general formula for pull-backs, \eg
\beq
P[G_{ab}]=G_{\mu\nu}{\prt x^\mu\over\prt\sig^a}{\prt x^\nu\over\prt\sig^b}
\labell{pullgen}
\eeq
reduces to that given in eq.~\reef{pull}.

As described in the introduction, the low-energy dynamics of
N coincident D-branes is described by a nonabelian U(N) gauge theory
\cite{bound}. Our conventions are such that
\beq
A_a=A_a^{(n)}T_n,
\qquad F_{ab}=\prt_a A_b-\prt_b A_a+i[A_a,A_b]
\labell{gauge}
\eeq
where $T_{n}$ are $\NN^2$ hermitian generators\footnote{There are certain
subtleties to describing D-branes on a compact space \cite{waticom,watirev},
but we will not have to consider these in the following.} 
with $\Tr(T_{n}\,T_m)=\NN\,\delta_{n m}$.
The gauge fields are accompanied by $9-p$ adjoint scalars $\Phi^i$
with
\beq
D_a\Phi^i=\prt_a\Phi^i+i[A_a,\Phi^i]\ .
\labell{scalder}
\eeq
Note that our conventions are such that both the gauge fields
and adjoint scalars have the dimensions of $length^{-1}$
--- hence the appearance of the string scale in $x^i=2\pi\ls^2\Phi^i$.

Now T-duality acts to change the dimension
of a D$p$-brane's world-volume \cite{Polchin2}. The two possibilities are:
{\it (i)} if a coordinate transverse to the D$p$-brane, \eg $y=x^{p+1}$, is
T-dualized, it becomes a D$(p+1)$-brane where $y$ is now the
extra world-volume direction; and {\it (ii)} if a world-volume coordinate 
on the D$p$-brane, \eg $y=x^{p}$, is
T-dualized, it becomes a D$(p-1)$-brane where $y$ is now an
extra transverse direction. In the process, the role of the
corresponding world-volume fields change as
\beq
(i)\ \Phi^{p+1}\,\rightarrow\, A_{p+1}\ ,
\qquad\qquad
(ii)\ A_p\,\rightarrow\,\Phi^p\ ,
\labell{rule1}
\eeq
while the remaining components of $A$ and scalars $\Phi$ are left
unchanged.

If we focus on case {\it (ii)}, the corresponding
field strengths appearing in the world-volume action, \reef{biact}
and \reef{csact}, transform accordingly
\beq
F_{ap}\ \longrightarrow\ D_a\Phi^p\ . \labell{Wrule1}
\eeq
We will show below that the modified interactions appear precisely
in the correct way to yield the pull-backs \reef{pull} to the reduced
world-volume. The pull-backs to the original D$p$-brane world-volume
will in general involve $D_p\Phi^i$. Now one would
apply a T-duality transformation along $x^p$ only when all fields are
independent of this coordinate. Hence if we were considering a
single D$p$-brane with abelian world-volume fields, we would have
\beq
D_p \Phi^i=\partial_p\Phi^i=0
\labell{simple}
\eeq
and so T-duality would not generate any new interactions from
the pull-backs. However, in the nonabelian case, we would only
require
\beq
D_p \Phi^i=\partial_p\Phi^i+i[A_p,\Phi^i]=i[A_p,\Phi^i]\ .
\labell{nonsimple}
\eeq
which need not vanish in general.\footnote{Here, one can think that the
scalars $\Phi^i$ are constant along $x^p$ up to a gauge transformation
\vskip -.4cm
\beq
\cL_v\Phi^i=v^aD_a \Phi^i=i[\Lambda,\Phi^i]
\labell{expl}
\eeq
\vskip -.2cm
\noindent where $v^a\prt_a=\prt_p$ is the Killing vector in the
$x^p$-direction, and the gauge parameter would be $\Lambda=v^aA_a$.
This point of view is analogous to the generalized analysis of T-duality for
spacetime fields in ref.~\cite{Tduo}.}
Hence T-duality would yield
\beq
D_p\Phi^i\ \longrightarrow\ i[\Phi^p,\Phi^i]\ . \labell{Wrule2}
\eeq
generating new world-volume interactions involving scalar
commutator terms. These interactions would not be included
if we simply introduced an overall gauge trace and substituted
nonabelian fields everywhere in eqs.~\reef{biact} and \reef{csact},
and hence the resulting actions are inconsistent with T-duality.
It is taking care to retain
all of the commutators, which distinguishes the present investigation
from previous discussions of T-duality in the context of D-branes
\cite{tfirst,tooltime}. 

With the conventions, the tension of an individual D$p$-brane,
\ie the coefficient appearing in the Born-Infeld term \reef{biact}, is
as usual \cite{Polchin2}
\beq
T_p={2\pi\over g(2\pi\ls)^{p+1}}\ .
\labell{tension}
\eeq
However, because of the ``unconventional'' normalization of the
RR fields noted above, the charge of an individual D$p$-brane,
\ie the coefficient appearing in the Chern-Simons action \reef{csact}, is
\beq
\mu_p={2\pi\over g(2\pi\ls)^{p+1}}\ .
\labell{charge}
\eeq
Hence we have $\mu_p=T_p$, which simplifies some of our calculations.
Finally, in the following discussion, it will be convenient to
define:
\beq
\l=2\pi\ls^2\ .
\labell{lambchop}
\eeq

\section{Born-Infeld Action} \labels{act}

In the previous section, we showed that merely extending the
abelian Born-Infeld action \reef{biact} with the substitution
of nonabelian fields and an overall gauge trace would miss certain
interactions involving commutators of the scalar fields. For example,
in a particular limit, the correct action for N coincident D$p$-branes
should reduce to $(p+1)$-dimensional U(N) super-Yang-Mills theory which
has a scalar potential proportional to $\Tr\,[\Phi^i,\Phi^j]^2$. This
potential would not appear in the naive construction suggested above.

Our approach will be to apply this construction, but only for the
D9-brane action to provide a starting point.
In this case, there are no world-volume scalar fields,
because the D-brane fills the entire space and
there are no transverse directions. Given the absence of any scalars,
making the naive construction will not omit the commutator terms described
above.\footnote{Of course, we could have also begun with the D8-brane
where again there can be no nontrivial commutators because there is a single
transverse scalar field.} Thus our nonabelian Born-Infeld action for
the D9-brane is
\beq
S_{BI}=-T_9 \int d^{10}\sigma\ \Tr\left(e^{-\phi}\sqrt{-det\left(
G_{ab}+B_{ab}+\l\,F_{ab}\right)}\right)\ .
\labell{biact9}
\eeq
Note that in the absence of any transverse directions, there is no
need to introduce a pull-back on $G_{ab}+B_{ab}$. 
Now our strategy is to apply a T-duality transformation, using
eqs.~(\ref{transformed}), \reef{dilaton} and (\ref{rule1}), on $9-p$ 
coordinates $x^i$ with $i=p+1,\ldots,9$ to
eq.~\reef{biact9} to produce the corresponding Born-Infeld action for a
D$p$-brane. Note that in eq.~\reef{biact9}, the background
fields are in general functions of all of the world-volume coordinates.
For the purposes of the T-duality transformation, we must consider the
special case that the fields are independent of the coordinates $x^i$.
This produces a D$p$-brane action functional of the background fields for
this special case, but we assume the extension to the general case
is simply to allow the background fields in the new action to be
functions of all of the spacetime coordinates. The latter entails
introducing a functional dependence on the nonabelian scalars
--- see eq.~\reef{slick} below --- which may seem a radical alteration,
but recall that string scattering amplitudes have already provided
evidence for this structure \cite{scatt,scatt2}. 

Focusing on the determinant in the D9-brane action, in the
notation of eq.~\reef{eee}, we have
\beq
D=det\left(E_{ab}+\lambda\,F_{ab}\right)
\labell{nine}
\eeq
and then applying the T-duality transformation rules
(\ref{transformed},\ref{rule1}) yields
\beq
\tilde{D}=det\pmatrix{E_{ab}-E_{ai}E^{ij}E_{jb}+\lambda\,F_{ab}&
E_{ak}E^{kj}+\lambda\,D_a\Phi^j\cr
-E^{ik}E_{kb}-\lambda\,D_b\Phi^i&E^{ij}+i\lambda\,[\Phi^i,\Phi^j]\cr}
\labell{transform}
\eeq
where as before, $E^{ij}$ denotes the inverse of $E_{ij}$.

Now it is instructive to consider this construction with the abelian
theory for which the commutators vanish and $D_a\Phi^i=\prt_a\Phi^i$.
In this case, the matrix inside the determinant above may be manipulated to
give
\beqa
\tilde{D}&=&det\pmatrix{E_{ab}-E_{ai}E^{ij}E_{jb}+\lambda\,F_{ab}&
E_{ak}E^{kj}+\lambda\,\prt_a\Phi^j\cr
\quad+(E_{ak}E^{kl}+\lambda\,\prt_a\Phi^l)(E_{lb}+\lambda\,E_{lm}
\prt_b\Phi^m)& \cr
0&E^{ij}\cr}\nonumber\\
&=&det\pmatrix{E_{ab}+\lambda\,\prt_a\Phi^kE_{kb}
+\lambda\,E_{ak}\prt_b\Phi^k&E_{ak}E^{kj}+\lambda\,\prt_a\Phi^j\cr
\quad+\lambda^2\,\prt_a\Phi^kE_{kl}\prt_b\Phi^l+\lambda\,F_{ab}& \cr
0&E^{ij}\cr}\nonumber\\
&=&det\left(P[E]_{ab}+\lambda\,F_{ab}\right)\,det(E^{ij})\ .
\labell{noncomt}
\eeqa
Hence the scalar derivative terms which were generated by T-duality
provide precisely necessary terms to yield the pull-back of $E_{ab}$
to the new reduced world-volume. Next by the transformation rule
\reef{dilaton}, the dilaton factor in eq.~\reef{biact9} is replaced by 
\beq
e^{-\phi} \rightarrow {e^{-\phi}\over\sqrt{E^{ij}}}
\labell{dilly}
\eeq
which provides precisely the necessary factor to cancel the
second determinant above when the latter appears under the square root
in the action. In the D9-brane action, the invariant directions, \ie $x^i$,
can be integrated out before the T-duality transformation
so that the overall prefactor becomes  $T_9\prod_{i=p+1}^9 (2\pi R_i)$.
Now taking into account the transformations of the radii, $R_i\rightarrow
\ls^2/R_i$, and of the string coupling in $T_9$, $g\rightarrow
g\ls^{9-p}/\prod_{i=p+1}^9R_i$, this leaves an overall factor of
precisely $T_p$ in front of the T-dual action \cite{Polchin2}.
Assembling all of these factors yields the T-dual action
\beq
\tilde{S}_{BI}=
-T_p \int d^{p+1}\sig\ \left(e^{-\phi}\sqrt{-det(P[G+B]_{ab}+
\l\,F_{ab})}\right)
\labell{biabel}
\eeq
which has precisely the desired form. Thus we have recovered the
result that in the abelian theory, the Born-Infeld action \reef{biact}
is compatible with T-duality \cite{tfirst}.

Unfortunately the full determinant \reef{transform}
including the commutator terms does not lead to such a simple
result as in eq.~\reef{noncomt}. In this case, it is useful to define 
\beq
Q^i{}_j\equiv\delta^i{}_j+i\lambda\,[\Phi^i,\Phi^k]\,E_{kj}\ .
\labell{extra}
\eeq
Then following manipulations as in eq.~\reef{noncomt}, one finds
\beq
\tilde{D}=det\left(P\left[E_{ab}
+E_{ai}(Q^{-1}-\delta)^{ij}E_{jb}\right]+\lambda\,F_{ab}\right)\,det(E^{ij})
\,det(Q^i{}_j)\ .
\labell{noncomt2}
\eeq
Note the second index on the expression $(Q^{-1}-\delta)^{ij}$ has
been raised using $E^{ij}$ (rather than $G^{ij}$). The remaining
ingredients of the calculation are precisely as above, and
the final T-dual action becomes
\beq
\tilde{S}_{BI}=-T_p \int d^{p+1}\sigma\,\Tr\left(e^{-\phi}\sqrt{-det\left(
P\left[E_{ab}+E_{ai}(Q^{-1}-\delta)^{ij}E_{jb}\right]+
\l\,F_{ab}\right)\,det(Q^i{}_j)}
\right)\ .
\labell{biactcom}
\eeq
An interesting features of this result is that it now contains
a product of two determinants.\footnote{This feature was noted for the
flat space action in ref.~\cite{yet2}.} It is the second determinant
which supplies the standard scalar potential in the flat space limit.
Setting $G_{\mu\nu}=\eta_{\mu\nu}$
and $B_{\mu\nu}=0$, one may expand this factor to find
\beq
\sqrt{det\,Q^i{}_j}=1-{\lambda^2\over4}[\Phi^i,\Phi^j]
[\Phi^i,\Phi^j]+\ldots
\labell{potent}
\eeq
Of course, this expansion also contains higher order potential terms, and
in general, more commutator terms arise in the first determinant factor
through the expression $(Q^{-1}-\delta)^{ij}$.

Now the Born-Infeld action is highly nonlinear, and so the
proposed action \reef{biactcom} is incomplete without a precise
prescription for how the gauge trace should be implemented.
To incorporate the nonabelian symmetry of the world-volume fields
Tseytlin\cite{yet} --- see also \cite{yet2} ---
has suggested that one should supplement
the usual Born-Infeld form (for a flat space background)
with a symmetrized trace over gauge indices. That is, to leading order
all commutators of the field strengths should be dropped.
Unfortunately this action does not seem to capture the
full physics of the infrared limit\cite{notyet}, and it appears that
in general nontrivial commutators must be included at sixth-order in
the field strength\cite{notyet2,notyet3}. On the other hand in configurations
with supersymmetry, this action seems to properly describe the physics
of the nonabelian Yang-Mills fields \cite{better}, and even provides
solutions of the full open-string equations of motion \cite{best}.

In the above action \reef{biactcom}, we will adopt the same
symmetrized trace prescription. That is, the trace is completely
symmetric between all nonabelian expressions of the form
$F_{ab}$, $D_a\Phi^i$ and $i[\Phi^i,\Phi^j]$. In part, this
prescription is adopted as a practical matter as these objects
were treated as commuting in all of the manipulations leading up
to eq.~\reef{biactcom}. Given the previous results \cite{notyet2,notyet3}
then,
we expect that in general this action will only incorporate the
correct interactions involving these objects at second and
fourth order. So for example, potential terms arising from the expansion of
$det(Q)$ in eq.~\reef{potent} may not be reliable for sixth and higher
orders in the commutators.
However, there is some hope that it will still properly
describe the physics of supersymmetric configurations to all orders.

This discussion has not addressed how the gauge trace should
be implemented for the nonabelian scalars appearing in the
functional dependence of the background fields. We will leave
this question until section \ref{cheque}.

\section{Chern-Simons Action} \labels{coup}

For the abelian theory of an individual D-brane, one can show that the
Chern-Simons action \reef{csact} is compatible with T-duality \cite{tfirst}.
In general the analysis is quite complicated because of the interplay of
the RR and the Neveu-Schwarz fields.\footnote{Recently there have
been some interesting suggestions on the transformation properties of
the RR fields for higher dimensional tori\cite{zoo}. This formalism
may simplify the analysis of the general case in the present discussion.}
For example, T-duality \reef{NSrule}
acting on the two-form $B$ in general introduces terms involving
$G_{\mu y}$, but these are all precisely cancelled by
the terms involving the metric appear in the transformation of RR potentials
\reef{RRrule}. The discussion, however, is greatly
simplified for the case of flat space, \ie $G_{\mu\nu}=\eta_{\mu\nu}$ and
$B_{\mu\nu}=0$. In this case, the T-duality transformations of the
RR potentials become \cite{Polchin2}
\beq
\tilde{C}^{(n)}_{\mu\cdots\nu\al y}=
C^{(n-1)}_{\mu\cdots\nu\al}\ ,\qquad
\tilde{C}^{(n)}_{\mu\cdots\nu\alpha\beta}=
C^{(n+1)}_{\mu\cdots\nu\alpha\beta y}\ ,
\labell{RRruls}
\eeq
while the action \reef{csact} reduces to
\beq
S_{CS}=\mu_p\int P\left[\sum C^{(n)}\right]e^{\l\,F}\ .
\labell{cssimp}
\eeq
Verifying the consistency of the form of this action with T-duality is
relatively straightforward, and we sketch the calculation here.
We consider the case where T-duality acts on a world-volume coordinate,
\eg $y=x^p$, so the D$p$-brane becomes a D$(p-1)$-brane. Now the
integrand in eq.~\reef{cssimp} is a form on the world-volume and so
the index $y$ will appear on precisely one tensor in each of the terms
in the sum. For the abelian theory, the pull-back involves ordinary
derivatives of the scalars, and so as $\prt_y\Phi^i=0$, the existing
pull-back terms are
unaffected by the T-duality transformation, \ie they already
coincide with the pull-back to the new reduced world-volume
from the transverse directions $x^i$ with $i=p+1,\ldots,9$.
Hence we may consider a generic term of the form
\beq
P[C^{(n)}]_{a_1\cdots a_n}\, F_{b_1c_1} \cdots F_{b_mc_m}
\labell{term}
\eeq
where again $y$ appears as one and only one of the indices.
If this index is carried by the RR potential, then the T-duality
transformation \reef{RRruls} simply removes this index leaving
\beq
P'[C^{(n-1)}]_{a_1\cdots a_{n-1}}\, F_{b_1c_1} \cdots F_{b_mc_m}
\labell{term1}
\eeq
where the prime on the pull-back indicates that it only involves
$\prt_a\Phi^i$ with $i=p+1,\ldots,9$. The other possibility
is that the $y$ index is carried by one of the gauge field strengths,
in which case eqs.~\reef{RRruls} and \reef{Wrule1} yield a contribution
of the form
\beq
m\,P'[C^{(n+1)}]_{a_1\cdots a_{n}y}\,
F_{b_1c_1} \cdots F_{b_{m-1}c_{m-1}}\, \prt_{b_m}\Phi^y\ .
\labell{term2}
\eeq
Hence these contributions complete the pull-back $P'$
which would have appeared in eq.~\reef{term1} if we had
begun with $n+2$ and $m-1$ in eq.~\reef{term}. This gives the
essence of the calculation. Checking the consistency
in detail requires
verification that a few factors and signs work out properly, and
that the prefactor in the T-dual action becomes $\mu_{p-1}$
(in the same way as $T_p$ is transformed in the previous section).
In the end, one does find that the action \reef{cssimp}
is compatible with T-duality \cite{tfirst}. The calculations become far more
involved for the case of general background fields, \ie arbitrary $G_{\mu\nu}$
and $B_{\mu\nu}$,  however the consistency of the full 
Chern-Simons action \reef{csact}
can still be verified on a case-by-case basis \cite{tfirst}.

Now we argued in section \ref{prelim}
that the form action in eqs.~\reef{csact} and \reef{cssimp}
fails to include certain interactions involving the commutators
of the nonabelian scalar fields. So let us simply present the
nonabelian generalization of the flat space Chern-Simons action
\reef{cssimp}:
\beq
S_{CS}=\mu_p\int \Tr\left(P\left[e^{i\l\,\hi_\Phi \hi_\Phi}\sum C^{(n)}\right]
e^{\l\,F}\right)\ .
\labell{cssimp1}
\eeq
Beyond the generalization of all of world-volume fields to their
nonabelian counterparts (including in the functional dependence of
the RR potentials) and the inclusion of an overall gauge trace, there
is the appearance of the exponential $\exp(i\l\,\hi_\Phi \hi_\Phi)$ inside
the pull-back. Here $\hi_\Phi$ denotes the interior product by $\Phi^i$
regarded as a vector in the transverse space. Acting on forms,
the interior product is an anticommuting operator of form degree --1,
\eg
\beqa
C^{(2)}&=&{1\over2} C^{(2)}_{\mu\nu}\,dx^\mu dx^\nu
\labell{interior}\\
\hi_vC^{(2)}&=&v^\mu\, C^{(2)}_{\mu\nu}\,dx^\nu
\nonumber\\
\hi_w\hi_vC^{(2)}&=&w^\nu v^\mu\, C^{(2)}_{\mu\nu}
=-\hi_v\hi_wC^{(2)}\ .
\nonumber
\eeqa
In particular, for ordinary vectors one has $(\hi_v)^2=0$.
The exponential makes a nontrivial contribution in eq.~\reef{cssimp1}
because of the nonabelian nature of the displacement vectors
$\Phi^i$. There one has
\beq
\hi_\Phi \hi_\Phi C^{(2)} = \Phi^j\Phi^i\,C^{(2)}_{ij}={1\over2}C^{(2)}_{ij}
\,[\Phi^j,\Phi^i]
\labell{nonabint}
\eeq
and so the exponential produces precisely the desired commutators.

Verifying the consistency of the action \reef{cssimp1} with T-duality
follows essentially the same argument as above for the abelian action
\reef{cssimp}. Again, we outline the calculation where
T-duality acts on a world-volume coordinate, $y=x^p$.
The only real difference is that the pull-back now involves 
gauge covariant derivatives of the scalars. Hence since
$D_y\Phi^i=i[A_y,\Phi^i]$ which in general is nonvanishing,
T-duality acting on the pull-back will now give rise to
new interactions.
Here, we consider a generic term of the form
\beq
(i\l)^l P[(\hi_\Phi \hi_\Phi)^lC^{(n+2l)}]_{a_1\cdots a_n}\, F_{b_1c_1}
\cdots F_{b_mc_m}
\labell{nterm}
\eeq
where again $y$ appears as one and only one of the indices.
First since the scalars above only carry indices $i=p+1,\ldots,9$, the
operator $(\hi_\Phi \hi_\Phi)^l$ will remain unaffected by the T-duality.
If the $y$ index is carried by the RR potential after the
pull-back, then the T-duality
transformation \reef{RRruls} simply removes this index leaving
\beq
{(i\l)^l\over l!}
 P'[(\hi_\hPhi \hi_\hPhi)^lC^{(n+2l-1)}]_{a_1\cdots a_{n-1}}
\, F_{b_1c_1}\cdots F_{b_mc_m}
\labell{nterm1}
\eeq
where the prime on the pull-back indicates that it only involves
$D_a\Phi^i$ with $i=p+1,\ldots,9$, and similarly $\hi_\hPhi$ indicates
that the displacement vector in the interior product only has
components in these same directions, \ie the original transverse space.
Another possibility is that $y$ appears as one of the indices $a_k$
in eq.~\reef{nterm}, but that it is actually carried by $D_y\Phi^i$
in the pull-back. In this case, the T-duality transformation yields
\beq
2{(i\l)^{l+1}\over l!}
\,P'[(\hi_\hPhi \hi_\hPhi)^l(\hi_{\vphantom{\hPhi}\Phi^y}\hi_\hPhi)C^{(n+2l+1)}
]_{a_1\cdots a_{n-1}}\, F_{b_1c_1}\cdots F_{b_mc_m}\ .
\labell{nterm11}
\eeq
The role of these terms is to complete the interior product operators
that would appear in a term given in eq.~\reef{nterm1} if one replaced
$l$ by $l+1$ in eq.~\reef{nterm}. That is, one can write
\beq
{(i\l)^{l+1}\over (l+1)!}(\hi_\Phi \hi_\Phi)^{l+1}=
{(i\l)^{l+1}\over (l+1)!}(\hi_\hPhi \hi_\hPhi)^{l+1}+2{(i\l)^{l+1}\over l!}
(\hi_\hPhi \hi_\hPhi)^l(\hi_{\vphantom{\hPhi}\Phi^y}\hi_\hPhi)
\labell{intell}
\eeq
when one realizes that $\Phi^y$ can at most appear in one of the interior
products since the combined operator is acting on a form.
The final possibility, as in the abelian case,
is that the $y$ index is carried by one of the gauge field strengths,
in which case eqs.~\reef{RRruls} and \reef{Wrule1} yield a contribution
of the form
\beq
m\,(i\l)^l P'[(\hi_\Phi \hi_\Phi)^lC^{(n+2l+1)}]_{a_1\cdots a_{n}y}\,
F_{b_1c_1} \cdots F_{b_{m-1}c_{m-1}}\, D_{b_m}\Phi^y\ .
\labell{nterm2}
\eeq
As before the role of these contributions is to complete the pull-back $P'$
which would have appeared in eq.~\reef{nterm1} if we had
begun with $n+2$ and $m-1$ in eq.~\reef{nterm}. One point to note
is that in collecting contributions to form interactions of the
form given in eq.~\reef{nterm} but now on the reduced world-volume
with $a=0,\ldots,p-1$, one will never find $\hi_{\Phi^y}$ and $D_a\Phi^y$
in the same term as they cannot both be contracted on the same RR form.
The above discussion describes the
core of the consistency calculations. Again we have left out
some of the details, and simply state that one will find
T-duality will transform the action \reef{cssimp1} for a D$p$-brane
to the D$(p-1)$-brane action with the same precise form.
The reader is invited to verify these details for herself.

In extending eq.~\reef{cssimp1} beyond the case of flat space, the
Chern-Simons action for the nonabelian world-volume theory becomes
\beq
S_{CS}=\mu_p\int \Tr\left(P\left[e^{i\l\,\hi_\Phi \hi_\Phi} (
\sum C^{(n)}\,e^B)\right]
e^{\l\,F}\right)\ .
\labell{csnon}
\eeq
It is clear that this action reduces to the expected form \reef{csact}
for the abelian theory of a single D$p$-brane. 
The previous discussion verifying consistency with T-duality of
the action \reef{cssimp1} for flat space
generalizes here in a straightforward way for more general
backgrounds, as long as we impose the restriction that $G_{\mu y}=0$
which keeps the transformations \reef{RRrule} of the RR potentials
relatively simple. One fact that is revealed by such an analysis is
that the interior products in eq.~\reef{csnon} must act on both
the Neveu-Schwarz two-form and the RR potentials. For example,
one would find such interior products in eq.~\reef{nterm11} if the 
general transformation of the RR potential was applied in the previous
discussion. Calculations to verify that eq.~\reef{csnon} is consistent
with T-duality $G_{\mu\nu}$ is completely general become much more
involved, however, we have verified the consistency of this action \reef{csnon}
in certain specific cases, \eg considering the D2-brane with arbitrary
$G_{\mu\nu}$ and $B_{\mu\nu}$, but $C^{(n)}=0$ for $n\ge5$. 

Let us make a few remarks on the interpretation of eq.~\reef{csnon}.
First, note that the integrand is to be evaluated 
by considering the expression in each set of brackets in turn,
from the innermost around $\sum C^{(n)}e^B$ to the outermost for the
gauge trace. In particular, this means that first $\sum C^{(n)}e^B$
is expanded as a sum of forms in the ten-dimensional spacetime,
and so only a finite set of terms in the exponential will contribute.
Then the exponential $\exp({i\l\,\hi_\Phi \hi_\Phi})$ acts on this sum
of forms, and so again only a finite set of terms in this second exponential
will contribute.

Second, just as in the previous section with the Born-Infeld action, the
proposed action \reef{csnon} is incomplete without a precise
prescription for how the gauge trace should be implemented.
Here, we will adopt the same symmetrized trace prescription
as in the Born-Infeld action. That is, the trace is completely
symmetric between all nonabelian expressions of the form
$F_{ab}$, $D_a\Phi^i$ and $i[\Phi^i,\Phi^j]$. Further here we must
extend this symmetrization to the individual background field
components, each of which is in general a functional of the nonabelian
scalars. This symmetrization must be adopted here as a practical
matter as all of these objects were treated as commuting in the calculations
which verified that eq.~\reef{csnon} was compatible with T-duality.
Given the complicated possible functional dependence on $\Phi^i$ which
could appear in the background fields, the above description of the
gauge trace is still not complete. We will leave the question of its
precise implementation to the following section.

It is well-known that a D$p$-brane couples not only to the RR potential with
form degree $n=p+1$ \cite{Polchin}, but also that it can couple to
the RR potentials with $n=p-1,p-3,\ldots$ through the additional interactions
with the two-form $B$ and world-volume field strength $F$ appearing in
the action \reef{csact} \cite{mike}. Here we see from 
the nonabelian action \reef{csnon} that a D$p$-brane can also couple
to the RR potentials with $n=p+3,p+5,\ldots$ through the additional
interactions involving commutators of the nonabelian scalars. To make these
couplings more explicit, consider the D0-brane action (for which $F$
vanishes):
\beqa
S_{CS}&=&\mu_0\int \Tr\left(P\left[C^{(1)}+
i\l\,\hi_\Phi \hi_\Phi\left(C^{(3)}+C^{(1)}B\right)
\vphantom{\l^4\over24}\right.\right.
\labell{cszero}\\
&&\qquad -{\l^2\over2}(\hi_\Phi \hi_\Phi)^2\left(C^{(5)}
+C^{(3)}B+{1\over2}C^{(1)}B^2\right)
\nonumber\\
&&\qquad -i{\l^3\over6}(\hi_\Phi \hi_\Phi)^3\left(
C^{(7)}+C^{(5)}B
+{1\over2}C^{(3)}B^2+{1\over6}C^{(1)}B^3\right)
\nonumber\\
&&\qquad \left.\left.+{\l^4\over24}(\hi_\Phi \hi_\Phi)^4\left(
C^{(9)}+C^{(7)}B+{1\over2}C^{(5)}B^2
+{1\over6}C^{(3)}B^3+{1\over24}C^{(1)}B^4\right)\right]\right)
\nonumber\\
&=&\mu_0\int dt\, \Tr\left(\vphantom{\l\over2}
C_t^{(1)}+\l\,C_i^{(1)}D_t\Phi^i
+i{\l\over2}(C_{tjk}^{(3)}\,[\Phi^k,\Phi^j]+\l\,C_{ijk}^{(3)}
\,D_t\Phi^i\,[\Phi^k,\Phi^j]) +\ldots \right)
\nonumber
\eeqa
where we assume that $\sigma^0=t$ in static gauge.
We see here that the nonabelian action \reef{csnon} is giving interactions
reminiscent of those appearing in matrix theory \cite{matrix,matbrane}.
This similarity is, of course, no accident, as we will see in the
next section by comparing to the couplings derived from matrix theory
by Taylor and van Raamsdonk \cite{wati1}.
For example, there is linear coupling to $C^{3}$, which
is the potential corresponding to D2-brane charge,
\beq
i\lambda\,\mu_0\int \Tr\, P\left[\hi_\Phi \hi_\Phi C^{(3)}\right]
=i{\l\over2}\mu_0\int dt\ \Tr \left(C_{tjk}^{(3)}\,[\Phi^k,\Phi^j]
+\lambda C^{(3)}_{ijk}\,D_t\Phi^k\,[\Phi^k,\Phi^j]
\right)
\labell{magic}
\eeq
Note that the first term on the right hand side has the form of a source
for D2-brane charge. This is essentially the interaction central
to the construction of D2-branes in matrix theory with the large N
limit \cite{matrix,matbrane}. Here, however, with finite N,
this term vanishes  upon taking the trace if
$C_{tjk}^{(3)}$ was simply a constant or a function of the
world-volume coordinate $t$. However, in general one should regard
these components of the RR three-form as functionals of $\Phi^i$.
Hence, while there would be no ``monopole'' coupling to D2-brane charge,
nontrivial expectation values of the scalars can give rise to couplings
to an infinite series of higher ``multipole'' moments. In addition,
we will discuss in section \ref{diel} that in a nontrivial background
$C^{(3)}$, this interaction gives rise to additional terms in the
potential for the scalars.

\section{Comparison with Matrix Theory} \labels{cheque}

Thus far, we have constructed a nonabelian world-volume action
for test D$p$-branes in general background fields. The action consists
of two parts, the Born-Infeld term \reef{biactcom} and the Chern-Simons
term \reef{csnon}. Our construction was guided by the simple principle
that the result should be consistent with the familiar rules of T-duality.
In this section, we would like to discuss how our results measure up
to some previous considerations of nonabelian D$p$-brane actions. This
exercise will also be useful in providing a precise prescription for
the implementation of the gauge trace in our action --- see the
comparison to matrix theory, below.

First, we will remark on the connection of our Born-Infeld action
\reef{biactcom} to Tseytlin's proposal \cite{yet} for a nonabelian Born-Infeld
action for ten-dimensional super-Yang-Mills theory. Tseytlin's
action is, of course, precisely recovered in
eq.~\reef{biactcom} with $p=9$, \ie eq.~\reef{biact9}
and a flat space background,
\ie $G_{\mu\nu}=\eta_{\mu\nu}$ and $B_{\mu\nu}=0$. This action \cite{yet}
correctly yields the $F^2$ and $F^4$ interactions expected for low
energy superstring theory\cite{arkad4},
however, it seems to require modifications \cite{notyet2,notyet3} at order
$F^6$. Introducing a general metric (but keeping
$B_{\mu\nu}=0$) in our D9-brane action \reef{biact9} simply covariantizes the
super-Yang-Mills action with minimal coupling to the metric, \ie the same
$F^2$ and $F^4$ (and all higher order)
interactions are recovered but now the index contractions
are made with the curved space metric. Of course, T-duality
requires the inclusion of $B$ in eq.~\reef{biact9}, since T-duality
\reef{NSrule} interchanges components of the metric and this two-form.
As usual, the factor of $\exp({-\phi})$ is required in the action as
it must reproduce interactions derived from superstring disk
amplitudes \cite{Polchin2}. Hence we are arguing that eq.~\reef{biact9},
our starting point in the construction of the Born-Infeld action
\reef{biactcom},
provides the minimal extension of Tseytlin's action \cite{yet} to include
nontrivial Neveu-Schwarz backgrounds. As mentioned above,
the connection with Tseytlin's action also points out the limitations
of eq.~\reef{biactcom},
namely that we must expect that there are corrections
at order six in $F_{ab}$, $D_a\Phi^i$ and/or $i[\Phi^i,\Phi^j]$.
Given that the precise form of the sixth order corrections to
Tseytlin's action are known\cite{notyet2}, presumably
those to eq.~\reef{biactcom} could be determined by covariantizing the
flat space corrections, multiplying by $\exp({-\phi})$, and
demanding consistency with T-duality.

Douglas \cite{moremike} observed that whatever their form, the
nonabelian world-volume actions should contain a single gauge trace,
as do both eqs.~\reef{biactcom} and \reef{csnon}. This observation
again stems from the fact that these actions are encoding the
low energy interactions derived from disk amplitudes in superstring
theory. Since the disk has a single boundary,
the single gauge trace arises from the standard open
string prescription of tracing over Chan-Paton factors on each
world-sheet boundary. In particular, this means that the
background (closed string) fields appearing in the action
cannot be functionals of the D-brane center-of-mass coordinate
$x^i={\l\over N}\Tr\,\Phi^i$. Hence, it is natural to assume that
the background fields are functionals of the nonabelian scalar
fields instead \cite{moremike}. One can sharpen this reasoning
by observing that the only difference in the
superstring amplitudes between the U(1) and the U(N) theories is that
the amplitudes in the latter case are multiplied by an additional trace
of Chan-Paton factors. Hence up to commutator ``corrections,''
the low energy interactions should be the same in both cases. Hence
since the background fields are functionals of the neutral U(1) scalars
in the abelian theory, they must be precisely the same functionals
of the adjoint scalars in the nonabelian theory, up to commutator
corrections. Similarly the pull-back was
constructed with neutral scalars in the abelian theory,
and so the adjoint scalars
play the same role in the nonabelian theory \cite{hull}.

While we have indicated that the background fields are
functionals of the nonabelian scalars, let us be more
precise on how this is actually implemented \cite{scatt}.
For example, consider the case of a
D$p$-brane propagating in a curved background $G^0_{\mu\nu}(x^\rho)$.
First we impose static gauge as described above eq.~\reef{pullgen},
and hence the spacetime coordinates are split into
world-volume coordinates, $x^a=\sigma^a$, and transverse coordinates,
$x^i$. Then the metric functional appearing in
the D-brane action would be given by a {nonabelian} Taylor expansion
\cite{scatt}
\beqa
G_{\mu\nu}&=&\exp\left[\l\Phi^i\,{\prt_{x^i}}\right]G^0_{\mu\nu}
(\sigma^a,x^i)|_{x^i=0}
\labell{slick}\\
&=&\sum_{n=0}^\infty {\l^n\over n!}\,\Phi^{i_1}\cdots\Phi^{i_n}\,
(\prt_{x^{i_1}}\cdots\prt_{x^{i_n}})G^0_{\mu\nu}
(\sigma^a,x^i)|_{x^i=0}\ .
\nonumber
\eeqa
The same nonabelian Taylor expansions arise in the context of
matrix theory, \eg \cite{watikabat,waticur}.
Since the partial derivatives above all commute, there does not
seem to be room for commutator corrections to this
Taylor expansion, \ie it naturally produces an expression symmetric in
all of the $\Phi^{i_m}$. The ambiguities come in the full nonlinear action
where the trace must be implemented
over $\Phi^i$'s from different background field components, as well
nonabelian terms such as $F_{ab}$.

We also mention that explicit calculations of superstring
amplitudes \cite{scatt,scatt2,Garousi} yield a detailed agreement with
the actions given in eqs.~\reef{biactcom} and \reef{csnon}. In particular,
disk amplitudes with one closed string state and two open strings
\cite{scatt} reveal terms to second order in the nonabelian Taylor
expansion \reef{slick}, as well as contributions of the adjoint scalars
to the pull-backs. Some disk amplitudes with one closed and
three open strings have been evaluated \cite{scatt2}, and here some
of the single commutator corrections can be seen. For example, one sees the
commutator arising from the interior products in eq.~\reef{magic}.

The main comparison which we wish make is between certain interactions
in eqs.~\reef{biactcom} and \reef{csnon}, and those derived from
matrix theory \cite{wati1}. Taylor and van Raamsdonk \cite{wati1}
have recently calculated all of the linear couplings of (massless
bosonic) closed string fields to D0-branes in type IIa string theory,
by transcribing the analogous couplings to D=11 supergravity fields
derived for matrix theory \cite{watikabat,waticur}
in the Sen-Seiberg limit \cite{nati}. The nonabelian
action proposed here contains analogous linear couplings (as well
as nonlinear couplings). A detailed comparison the linear couplings
derived by these the two approaches yields an precise agreement,
to the order that the calculations are expected to be valid, up
to two caveats, which will be discussed below.\footnote{A further
caveat is that certain ambiguous signs in the matrix model result
are actually fixed by comparing to the D-brane couplings \cite{watiprep}.}
I will illustrate these comparisons by providing the detailed expressions
for two cases, the linear couplings of the dilaton and
the RR five-form potential. 

Taylor and van Raamsdonk \cite{wati1} present the linear coupling
to the dilaton as
\beq
\int d\htx\,\sum_{n=0}^\infty {1\over n!}
\left(\prt_{x^{i_1}}\cdots\prt_{x^{i_n}}\phi\right)\, I_\phi^{(i_1\cdots i_n)}
\ ,
\labell{dilcoup}
\eeq
where $\prt_{x^{i_1}}\cdots\prt_{x^{i_n}}\phi(s)$ are the derivatives of
the background dilaton field evaluated along the D0 world-line, \ie
the c-number coefficients in the Taylor series for the background field
\reef{slick}. The various moments have the form
\beq
I_\phi^{(i_1\cdots i_n)}=\STr(I_\phi,X^{i_1},\ldots,X^{i_n})
\labell{dilcurr}
\eeq
where the $X^i$ are the nonabelian matrix displacements completing the
Taylor expansion \reef{slick} of the background field. In eq.~\reef{dilcurr},
we have dropped an additional term including extra fermionic
contributions to the higher multipole moments --- as we have not included
the world-volume fermions in any of our calculations, we can make no
comparison for these terms.
As well as the nonabelian displacements, these currents are constructed
with $I_\phi$, the zeroth moment of the background field coupling.
The latter is an expression constructed from $D_\htx X^i$ and $i[X^i,X^j]$,
as well as $\Theta$ and $[X^i,\Theta]$ --- these fermionic terms will again
be dropped for the purposes of the present comparisons. The symmetrized
trace indicated by $\STr(\cdots)$ indicates that the gauge trace takes a
symmetrized average over all orderings of factors of $D_\htx X^i$ and
$i[X^i,X^j]$, and the individual nonabelian displacements $X^i$,
appearing in the higher moments.
This form of the coupling generalizes to all of the background fields,
and so the essential step is now to identify the form of the
zeroth moment currents.

{}From the expressions provided by Taylor and van Raamsdonk \cite{wati1}, one
determines the zeroth moment of the dilaton coupling to be
\beqa
I_\phi&=&{1\over R}\left(1 - {1\over2} D_\htx X^iD_\htx X^i-
{1\over 4} [X^i,X^j][X^i,X^j] -{1\over8}(D_\htx X^iD_\htx X^i)^2
\right.\labell{zeromom}\\
&&\quad\quad
-{1\over8}[X^i,X^j][X^j,X^k][X^k,X^l][X^l,X^i]
+{1\over32} ([X^i,X^j][X^j,X^i])^2
\nonumber\\
&&\quad\quad\left.
-{1\over8}[X^i,X^j][X^j,X^i]D_\htx X^kD_\htx X^k
+{1\over2}D_\htx X^i[X^i,X^j][X^j,X^k]D_\htx X^k\right)+O_c(v^4)+O(v^6)
\nonumber
\eeqa
where $R$ is the compactification
radius of the tenth spatial dimension in M-theory.
This expression gives the leading
order contributions in an expansion in small $D_\htx X^i$ and $i[X^i,X^j]$.
The $O(v^6)$ indicates that there will be new contributions at sixth order
in these expressions, while the $O_c(v^4)$ indicates that there
could also be additional commutator corrections at fourth order in
the expansion. Here we have again dropped all of the fermion
contributions calculated in ref.~\cite{wati1}.

Now let us compare to the linear dilaton coupling that arises in D-brane
action derived here. Expanding the Born-Infeld term \reef{biactcom}
around flat space keeping only the term linear in the dilaton yields
\beq
T_0\int dt\,\Tr\left[\phi(t,\Phi)\,\sqrt{(1-\l^2D_t\Phi^i\,
Q^{-1}_{ij}D_t\Phi^j)\,det Q_{ij}}\right]
\labell{dualdil}
\eeq
where from eq.~\reef{extra},
\beq
Q_{ij}=\delta_{ij}+i\l [\Phi^i,\Phi^j]
\labell{flatq}
\eeq
in flat space. We see that this can be interpreted precisely as the
general form given in eqs.~\reef{dilcoup} and \reef{dilcurr}. There is
the background field $\phi(t,\Phi)$ which implicitly contains a
nonabelian Taylor expansion, as in eq.~\reef{slick}. This multiplies
an expression involving $D_t\Phi$ and $i[\Phi^i,\Phi^j]$, \ie the zeroth
moment. Of course, all of the linear couplings derived from the D-brane
action will have this same general form, and so there is agreement between
two calculations at this general level. One
must make a detailed comparison of the zeroth moments then, to determine
if there is in fact a precise agreement between the two calculations.
In the present case, it is straightforward to show that the expansion
of the square root in eq.~\reef{dualdil} in fact yields precisely
the dilaton zeroth moment in eq.~\reef{zeromom}. Rather than
repeat the lengthy formula above, let us describe the expansion by
saying that it yields: (i) at zero order, simply 1, as in eq.~\reef{zeromom},
(ii) at second order, the standard low energy kinetic and potential terms
for the nonabelian scalar (up to an overall sign), as in eq.~\reef{zeromom},
and (iii) at fourth
order, the dimensional reduction of the stringy corrections appearing
in the d=10 SYM theory at fourth order in field strengths
\cite{arkad4,yet}, again as in eq.~\reef{zeromom} --- that the terms above
provide the
desired dimensionally reduced is manifest in the notation of ref.~\cite{wati1}.
We should also comment that all of the quantities in the calculations
of ref.~\cite{wati1} are dimensionless, \ie they have set $\l=1$. To restore
the standard engineering dimensions of our notation, one would
set: $X^i\rightarrow \sqrt{\l}\Phi^i,$ $\prt_{x^{i}}
\rightarrow \sqrt{\l}\prt_{x^{i}},$ $\htx \rightarrow t/\sqrt{\l},$
and $R\rightarrow{R}/\sqrt{\l}.$ Then the agreement between
eqs.~\reef{dilcoup} and \reef{dualdil} is completed
by matching $R^{-1}=T_0$, which agrees with the standard
duality between type IIa superstring theory and D=11 M-theory
where $R=g\ls$ \cite{parad,matrix}. In fact, the D-brane calculation
indicates that in fact there are no additional commutator corrections
at fourth order in the zeroth moment of the dilaton coupling \reef{zeromom}.
Further the expansion of the D-brane action could be carried out
to the next order, where we know there are extra commutator corrections
but these have been explicitly calculated \cite{notyet3}. Hence in
principle, one could use these results to calculate the $O(v^6)$
corrections to eq.~\reef{zeromom}. Of course, this is only the next
term in an infinite series, and so there would still be unknown
terms at $O(v^8)$.

At this point, we come to the first of the aforementioned caveats. 
In the D-brane calculation, the
higher moments of the dilaton coupling which are
implicit in eq.~\reef{dualdil} are actually ambiguous,
since we have as yet not provided a specific prescription for implementing
the gauge trace. On the other hand, there is no such ambiguity
in the matrix theory calculation which comes with a specific symmetrized trace
prescription. Hence matrix theory guides us to the appropriate
prescription: the gauge trace takes a
symmetrized average over all orderings of $F_{ab}$, $D_a\Phi^i$,
$i[\Phi^i,\Phi^j]$, and as well the individual $\Phi^i$ appearing in the
functional dependence of the background fields. We adopt this as
our trace prescription in both the Born-Infeld \reef{biactcom}
and Chern-Simons \reef{csnon} actions. This will ensure the
agreement of all of higher moments in the linear couplings
to those calculated from matrix theory \cite{wati1}. Note
that we are proposing that this prescription should be extended
to all of the nonlinear
couplings with background fields as well. This 
maximally symmetrized trace prescription also agrees with the
requirements that we outlined for the trace in the previous
two sections --- namely, it should be symmetric in $F_{ab}$, $D_a\Phi^i$
and $i[\Phi^i,\Phi^j]$, following ref.~\cite{yet}, and also that the
individual components of background fields should commute under the
trace.

Now let us consider the linear coupling of the RR five-form, which
Taylor and van Raamsdonk \cite{wati1} write as
\beq
\int d\htx\,\sum_{n=0}^\infty {1\over n!}
\left(\prt_{x^{i_1}}\cdots\prt_{x^{i_n}}
\tilde{C}^{(3)}_{\mu\nu\l\rho\sig}\right)
I_4^{\mu\nu\l\rho\sig(i_1\cdots i_n)}
\ ,
\labell{coup5}
\eeq
where their notation indicates that this potential generates the
field strength that is dual to that of the RR three-form.
Now the various moments again have the form
\beq
I_4^{\mu\nu\l\rho\sig(i_1\cdots i_n)}=\STr(I_4^{\mu\nu\l\rho\sig},
X^{i_1},\ldots,X^{i_n})\ .
\labell{curr5}
\eeq
As the zeroth moment now carries an antisymmetric set of spacetime indices
there are two distinct cases depending on whether or not one of the
indices coincides with the world-volume direction $\htx$. Taylor and
van Raamsdonk \cite{wati1} derived that
\beqa
I_4^{\htx ijkl}&=&-{3\over2R}\left([X^{[i},X^j][X^k,X^{l]}]+O(v^4)\,
\right)\ ,\labell{zeros4}\\
I_4^{ijklm}&=&6M^{+ijklm}-{15\over2R}\left(D_\htx X^{[i}[X^j,X^k][X^l,X^{m]}]
+O(v^5)\,\right)\ .\labell{zeroi4}
\eeqa
The $[\cdots]$ enclosing the indices indicates that the expressions
are completely antisymmetrized, and
as usual, all of the fermion contributions have been ignored.
In the D-brane action \reef{cszero}, the linear coupling to this
RR potential is
\beqa
&&
-{\l^2\mu_0\over2}\int dt\,\Tr\,P\left[(\hi_\Phi\hi_\Phi)^2C^{(5)}\right]
\labell{dcoup5}\\
&&\qquad=
-{\l^2\mu_0\over8}\int dt\,\Tr\left[ C^{(5)}_{tijkl}
[\Phi^{[i},\Phi^j][\Phi^k,\Phi^{l]}]+\l C^{(5)}_{ijklm}D_t\Phi^{[i}
[\Phi^j,\Phi^k][\Phi^l,\Phi^{m]}]\,\right]\ .
\nonumber
\eeqa
Comparing these expressions, one finds that there is precise
agreement between the explicit couplings
as before provided that the potentials in the two
calculations are related as: $C^{(5)}=60\,\tilde{C}^{(3)}$.
Further the D-brane results suggest that there are no higher
order corrections in eqs.~\reef{zeros4} and \reef{zeroi4},
or to the zeroth moments
of any of the RR potentials calculated in ref.~\cite{wati1}.

At this point, the reader should think that we have concluded too quickly
that there is a precise agreement in the previous calculation, as
we have not commented on the contribution $M^{+ijklm}$ in eq.~\reef{zeroi4}.
This brings us to the second of the caveats that we mentioned at the
outset above. Above, of course, we only find a precise agreement if
$M^{+ijklm}=0$. In matrix theory, $M^{+ijklm}$ is the transverse
5-brane charge, and determining this operator remains a famous unresolved
puzzle. So here through the D-brane action,
string theory is making a definite prediction as to the nature of this
operator, namely, $M^{+ijklm}=0$. This result is also required for the
agreement of other couplings, as well. The matrix theory calculation
\cite{wati1}
includes couplings to the six-form potential which would be the dual
of the Neveu-Schwarz two form. The zeroth moments in this case are
given by $M^{+ijklm}$ and $M^{ijklmn}$, both of which remain undetermined
in matrix theory. In the D-brane action, \reef{biactcom} and \reef{csnon},
there are simply no such couplings, and so agreement requires
both $M^{+ijklm}=0$ and $M^{ijklmn}=0$. Actually, here
and above, we have only really shown that what vanishes is
the part of these operators which is independent of the world-volume
fermions. However, it seems clear that introducing the fermions
in a supersymmetric action will not create any new couplings
to $\tilde{B}_{\mu\nu\l\rho\sig\tau}$.

\section{Dielectric-Branes} \labels{diel}

In this section, we begin an investigation the physical
effects arising from the new nonabelian interactions
in the full world-volume action,
\ie the sum of eqs.~\reef{biactcom} and \reef{csnon}.
In particular, here we will focus on the potential
for the nonabelian scalars. To begin,
consider this potential for D$p$-branes in a flat space background, \ie
$G_{\mu\nu}=\eta_{\mu\nu}$ with all other fields vanishing.
In this case, the entire scalar potential originates in the Born-Infeld
term \reef{biactcom} as
\beq
V=T_p\,\Tr\sqrt{det(Q^i{}_j)}= NT_p+{T_p\l^2\over4} 
\Tr([\Phi^i,\Phi^j]\,[\Phi^j,\Phi^i])+\ldots
\labell{trivpot}
\eeq
At low energies, the discussion focuses on the leading nontrivial
contribution above
\beq
{V}_{SYM}=-{T_p\l^2\over4} \Tr([\Phi^i,\Phi^j]\,[\Phi^i,\Phi^j])\ ,
\labell{trivpot1}
\eeq
which corresponds to the potential for ten-dimensional U(N)
super-Yang-Mills theory reduced to $p+1$ dimensions.
A nontrivial set of extrema of this potential (and the full expression
in eq.~\reef{trivpot}) is given by taking the $9-p$ scalars as constant
commuting matrices, \ie 
\beq
[\Phi^i,\Phi^j]=0
\labell{commat}\eeq
for all $i$ and $j$. Since they are commuting, the $\Phi^i$ may
be simultaneously diagonalized to give
\beq
\Phi^i=\pmatrix{x^i_1& 0& 0& \ddots\cr
                 0 &x^i_2& \ddots & 0\cr
		  0&\ddots &\ddots& 0 \cr
		 \ddots &0 &0 &x^i_\NN\cr}\ .
\labell{dmat}
\eeq
Here,
the eigenvalues are interpreted as the separated positions in the transverse
space of N fundamental D$p$-branes --- see, for example,
ref.~\cite{watirev}. This solution reflects the fact that a system of
N parallel D$p$-branes is supersymmetric, and they can rest in
static equilibrium with arbitrary separations in the transverse space
\cite{Polchin2}.

We have seen that going from flat space to a general background
introduces a vast number of new nonderivative scalar interactions
in the world-volume action. These
arise from the functional dependence
of the background fields on the scalars, and the appearance of extra
nonabelian commutators. Various physical effects governed by both of these
types of interactions have previously been considered. For example,
the former interactions are important in determining the dynamics of test
D$p$-branes in nontrivial backgrounds, \eg \cite{arkad2}.
In the context of matrix theory, the role of the second set of interactions
that was emphasized was to allow
D$p$-branes supporting noncommuting configurations of the scalars
to act a sources for RR potentials with form degree
greater than $p+1$, \eg \cite{wati1,matrix,matbrane,waticur}.

Here we want to consider another interesting physical effect
--- that is
the D-brane analog of the dielectric effect in ordinary electromagnetism.
When D$p$-branes are placed  in a nontrivial background
field for which the D$p$-branes would normally be regarded as neutral,
\eg nontrivial $F^{(n)}$ with $n>p+2$, new terms will be induced
in the scalar potential, and generically one should expect that there
will be new extrema beyond those found in flat space, \ie eq.~\reef{commat}.
In particular, there can be nontrivial extrema
with noncommuting expectation values of the $\Phi^i$, \eg with
$\Tr\Phi^i=0$ but $\Tr(\Phi^i)^2\ne0$. This would correspond to
the external field ``polarizing'' the D$p$-branes to expand
into a (higher dimensional)
noncommutative world-volume geometry. This is the analog of the familiar
electromagnetic process where an external field may induce
a separation of charges in neutral materials. In the latter, the
polarized material will carry then an electric dipole and possibly
higher multipoles. The D-brane analog of the latter is that
when the
world-volume theory is at a noncommutative extremum, the
nontrivial expectation values of the scalars will cause the
D$p$-branes to act as a source for new background fields.
To make these ideas explicit, we will illustrate the
process with a simple example below. Below, we consider N D0-branes
in a constant background RR field $F^{(4)}$, \ie the field strength
associated with D2-brane charge. We will find that the D0-branes expand into
a noncommutative configuration which represents the spherical bound state
of a D2-brane and N D0-branes.

\noindent{\bf \ref{diel}.1 Dielectric D0-branes}

Consider a collection of N D0-branes in a constant background
RR four-form field strength
\beq
F^{(4)}_{tijk}=\left\lbrace\matrix{-2f \vareps_{ijk}&{\rm for}\ i,j,k\in
\lbrace 1,2,3\rbrace\cr
0&{\rm otherwise}\cr}
\right.
\labell{backg}
\eeq
Note that $f$ carries the dimensions $length^{-1}$, and we
are assuming $\sigma^0=t$, with the static gauge condition.
Since $F^{(4)}=dC^{(3)}$, in order to construct the scalar potential,
we must consider the coupling of the D0-branes
to the  RR three-form potential,
\beqa
&&i\lambda\,\mu_0\int \Tr\, P\left[\hi_\Phi \hi_\Phi C^{(3)}\right]=
i\lambda\,\mu_0\int dt \Tr\left[\Phi^j\Phi^i\left(
C^{(3)}_{ijt}(\Phi,t)+\lambda
C^{(3)}_{ijk}(\Phi,t)\,D_t\Phi^k\right)\right]
\labell{interact}\\
&&\qquad\quad
=i\lambda\,\mu_0\int dt \Tr\left[\Phi^j\Phi^i\left(
C^{(3)}_{ijt}(t)+\lambda \Phi^k\prt_kC^{(3)}_{ijt}(t)+{\lambda^2\over2}
\Phi^l\Phi^k\prt_l\prt_kC^{(3)}_{ijt}(t)+\ldots\right.\right.
\nonumber\\
&&\qquad\qquad\qquad\qquad\left.\left.
+\lambda C^{(3)}_{ijk}(t)\,D_t\Phi^k+\lambda^2\Phi^l\prt_l
C^{(3)}_{ijk}(t)\,D_t\Phi^k+\ldots\right)\right]
\nonumber
\eeqa
where in the second line, we are explicitly introducing the
nonabelian Taylor series expansion \reef{slick} of the RR potential.
As noted below eq.~\reef{magic}, the quadratic term containing
$C^{(3)}_{ijt}(t)$ vanishes. Focusing on the terms that are cubic
in the scalar fields, we have
\beqa
&&i\lambda^2\mu_0\int dt\,\Tr\left(\Phi^j\Phi^i\left[
\Phi^k\prt_kC^{(3)}_{ijt}(t)+C^{(3)}_{ijk}(t)\,D_t\Phi^k\right]\right)
\labell{interact3}\\
&&\qquad={i\over3}\lambda^2\mu_0\int dt\,\Tr\left(\Phi^i\Phi^j\Phi^k\right)
F^{(4)}_{tijk}(t)\ .
\nonumber
\eeqa
Here, the second term in the first line was integrated by parts to
produce the final expression. This final form might have been anticipated
since one should expect that the induced potential can only depend on
gauge invariant expressions of the background field. Given that 
for simplicity we are considering a constant background $F^{(4)}$,
this interaction \reef{interact3} is the only term that need be considered.
All of the higher order terms implicit in eq.~\reef{interact} will
give rise to potential terms depending on derivatives of the four-form
field strength, and hence vanish. Combining eq.~\reef{interact3} with
the leading order Born-Infeld potential \reef{trivpot1} yields the
scalar potential of interest for the present problem
\beq
V(\Phi)=-{\lambda^2T_0\over4}\Tr([\Phi^i,\Phi^j]^2)
-{i\over3}\lambda^2\mu_0\Tr\left(\Phi^i\Phi^j\Phi^k\right)
F^{(4)}_{tijk}(t)\ .
\labell{potential}
\eeq

Substituting the background field \reef{backg} and demanding 
$\delta V(\Phi)/\delta\Phi^i=0$ yields the equation for extrema
\beq
0=[[\Phi^i,\Phi^j],\Phi^j]+{i}\,f\vareps_{ijk}[\Phi^j,\Phi^k]\ .
\labell{eqmot}
\eeq
Note that commuting matrices \reef{commat} still solve this
equation, and at this extremum describing separated D0-branes,
the value of the potential is simply $V_0=0$.

As an ansatz for a noncommuting solution, let us consider
constant scalars satisfying
\beq
[\Phi^i,\Phi^j]=2i\hR\,\vareps_{ijk}\Phi^k\ .
\labell{ansatz}
\eeq
Substituting this ansatz into eq.~\reef{eqmot} yields a solution for
\beq
\hR={f/2}\ .
\labell{solu}
\eeq
Hence we have nontrivial solutions of eq.~\reef{eqmot}
\beq
\Phi^i={f\over2}\,\alpha^i
\labell{solu1}
\eeq
where $\alpha^i$ are any N$\times$N matrix representation of
the SU(2) algebra
\beq
[\al^i,\al^j]=2i\,\vareps_{ijk}\,\al^k\ .
\labell{su2}
\eeq
For the moment, let us focus on the irreducible representation for
which one finds
\beq
\Tr[(\al_{\ssc \NN}^i)^2]={\NN\over3}(\NN^2-1) \quad{\rm for}\ i=1,2,3.
\labell{trace}
\eeq
Now evaluating the value of the potential \reef{potential}
for this new solution yields
\beq
V_\NN=-{T_0\l^2f^2\over6}\sum_{i=1}^3\Tr[(\Phi^i)^2]
=-{\pi^2\ls^3f^4\over6g}\NN(\NN^2-1)\ .
\labell{evpot}
\eeq
Hence the noncommutative solution solution has lower energy
than the solution of commuting matrices, and so the latter
configuration separated D0-branes is unstable towards
condensing out into this noncommutative solution.

Of course, the irreducible N$\times$N representation provides
only one solution for eq.~\reef{solu1}. One can compose
lower dimensional representations (including the trivial
1$\times$1 representation, 0) in a direct sum to produce
an alternative N$\times$N representation in eq.~\reef{solu1}.
However, such reducible representations would yield
$\Tr[(\al^i)^2]$ which is less than that for the irreducible
representation \reef{trace}. Hence, for the reducible
representations, one always finds
\beq
V_\NN<V_r\le V_0=0\ .
\labell{inbetween}
\eeq
Hence these noncommutative configurations based on reducible
representations would correspond to intermediate unstable
extrema of the potential \reef{potential}, and it appears that
the irreducible representation describes the ground state of
the system (and we refer to it as such in the following).

Given this ground state,
one should note that since both terms in the equations of
motion involve commutators, one can still modify the solution
by
\beq
\Phi^i={f\over2}\,\alpha_{\ssc \NN}^i + x^i \,I_{\ssc \NN}
\labell{solu2}
\eeq
where $I_{\ssc \NN}$ corresponds to the N$\times$N identity matrix.
This modification corresponds to shifting the center of mass
of the D0-branes to
\beq
{1\over \NN}\Tr(\Phi^i)=x^i
\labell{centermass}
\eeq
and of course, the value of the potential remains unchanged as
in eq.~\reef{evpot}. For the purposes of the following discussion
we will assume that $x^i=0$ without loss of
generality.\footnote{If $\alpha^i$ is reducible, one may make similarly
independent shifts for each of the irreducible representations in the
direct sum.}

Geometrically, one can recognize the algebra \reef{ansatz}
as that corresponding to the noncommutative or fuzzy two-sphere
\cite{fuzzball,fuzz}. In the context of matrix theory, this noncommutative
geometry was discussed in ref.~\cite{wati2}. The extent of
the noncommutative world-volume can be measured as
\beq
R=\l\left(\sum_{i=1}^3\Tr[(\Phi^i)^2]/\NN\right)^{1/2}=
\pi\ls^2f\NN\sqrt{1-{1\over \NN^2}}
\labell{radius}
\eeq
for the ground state solution.
For later convenience, we let us define the radius
$R_0\equiv\pi\ls^2f\NN$, which for large values of N gives
essentially the physical size of the noncommutative geometry.
One may infer from the matrix model construction of Kabat
and Taylor \cite{wati2} that the noncommutative solution actually
represents a spherical D2-brane with N D0-branes bound to it.
While such a configuration carries no net D2-brane charge, there
would be a ``dipole'' coupling due to the seperation of oppositely
oriented surface elements of membrane. The precise form of this
coupling can be calculated by substituting the noncommutative
scalar solution \reef{solu1} into the world-volume interaction
\reef{interact3}. Using the ground state solution, the result can
be written as
\beq
-{R_0^3\over3\pi g\ls^3}\left(1-{1\over \NN^2}\right)\int dt\,F^{(4)}_{t123}\ .
\labell{dipole}
\eeq
In fact, this is only the leading source term that would be generated
by substituting the noncommutative solution into the full expansion
in eq.~\reef{interact} --- that is, there will also be a series of
higher order ``multipole'' couplings, as well.

\noindent{\bf \ref{diel}.2 Dual D2-branes}

In the ground state solution above, the D0-branes condense out in
a noncommutative configuration, which we argued represents a bound
state of a spherical D2-brane and N D0-branes. It is interesting to
investigate to what extent these configurations can be matched in
the dual formulation of the same system. That is,
an abelian three-dimensional world-volume theory of a D2-brane
in which a flux of the U(1) gauge field strength
represents the N bound D0-branes. To parallel the same physics
as above, we consider this test D2-brane system in a constant
background RR four-form field strength \reef{backg}, and look
for a stable static configuration resembling that above.

To simplify the calculations, we write the flat space metric using
spherical polar coordinates to replace $x^{1,2,3}$, \ie
\beq
ds^2=-dt^2+dr^2+r^2\left(d\theta^2+\sin^2\theta\,d\phi^2\right)
+\sum_{i=4}^9(dx^i)^2\ .
\labell{flat}
\eeq
In static gauge, we choose the world-volume coordinates on the
D2-brane to be $\sigma^0=t,\sigma^1=\theta,\sigma^3=\phi$. We will look
for a static solution of the form 
\beq
r=R
\labell{RRRR}
\eeq
(and $x^i=0$ for $i=4,\ldots,9$).
The background four-form \reef{backg} should also be adapted to the
spherical polar coordinates
\beq
F^{(4)}_{tr\theta\phi}=-2fr^2\,\sin\theta\ \ 
\longrightarrow\ \ C^{(3)}_{t\theta\phi}={2\over3}fr^3\,\sin\theta
\labell{backg1}
\eeq
where we have made a convenient gauge choice for three-form potential.
The last ingredient required before proceeding to the action is the
U(1) field strength describing the N bound D0-branes,
\beq
F_{\theta\phi}={N\over 2}\sin\theta\ .
\labell{flux}
\eeq
One can confirm the normalization of this flux by considering the
induced coupling to the RR one-form in the Chern-Simons action
\reef{csact}. A simple calculation shows that
\beq
\l\mu_2\int C^{(1)}\,F = \NN\mu_0\int\! dt\,C^{(1)}_t
\labell{confirm}
\eeq
\ie the (monopole) coupling to N fundamental D0-branes is reproduced.

With the above results in hand, we turn to the action which in the
present case reduces to
\beq
S=-T_2\int \!dt\,d\theta\, d\phi \sqrt{-det(P[G]_{ab}+\l F_{ab})}
+\mu_2\int P[C^{(3)}]\ .
\labell{dualact}
\eeq
Note that for these calculations, we have a single D2-brane
and so the world-volume theory is abelian. Eq.~\reef{dualact}
fixes the Lagrangian density ${\cal L}$, and so for the static trial
solution \reef{RRRR}, the potential energy is given by
\beqa
V(R)&=&-\int\! d\theta\, d\phi\, {\cal L}
\labell{dualpot}\\
&=&4\pi T_2\left(\sqrt{R^4+{\l^2\NN^2\over 4}}-{2f\over3}R^3\right)
\nonumber\\
&=&\NN T_0 +{2T_0\over\l^2\NN}R^4+\ldots-{4T_0\over3\l}fR^3\ .
\nonumber
\eeqa
In the last line, we have expanded the square root for $2R^2/\l\NN<<1$,
only keeping the first two terms in the expansion explicitly, and we
have substituted $T_0=2\pi\l T_2$ everywhere.

The first thing that we note is that the constant term in the potential
energy corresponds precisely to the rest energy of N D0-branes. Thus this
precisely matches the same term in the nonabelian D0-brane calculation in the
previous section --- this is the ``trivial'' constant term in
eq.~\reef{trivpot}, which was then overlooked in eq.~\reef{potential}.
Keeping only the leading order terms given in eq.~\reef{dualpot},
there are two extrema:
\beqa
{\rm a)}&&R=0 \qquad\qquad\qquad{\rm with}\ V-\NN T_0=0\ ,
\labell{extrema}\\
{\rm b}&&R=R_0=\pi\ls^2f\NN\quad{\rm with}\ V-\NN T_0=-{\pi^2\ls^3f^4\over6g}
\NN^3\ .
\nonumber
\eeqa
The first extrema is an unstable inflection point in the potential,
while the second is a stable spherical configuration. Comparing the
equilibrium radius with eq.~\reef{radius}, we see that this calculation
reproduces the noncommutative ground state radius up to $1/\NN^2$ corrections.
Similarly the shift in the potential energy reproduces the result for
the D0-brane result \reef{evpot}, again up to $1/\NN^2$ corrections.

Finally, just as in the previous section, this spherical D2-brane
configuration carries no net D2-brane charge. However, because the
membrane is supported at a finite radius, there is a finite
dipole coupling:
\beq
\mu_2\int P[C^{(3)}]=
-{R_0^3\over3\pi g\ls^3}\int dt\,F^{(4)}_{t123}\ +\ \ldots
\labell{dualdipole}
\eeq
where we have only retained the leading term, and expressed the
result in Cartesian coordinates, to compare to eq.~\reef{dipole}.
Once again, the two calculations agree up to $1/\NN^2$ corrections.
It would be interesting to compare the higher order multipole
couplings in the two different frameworks.

\section{Discussion} \labels{conc}

Using T-duality, we have provided a straightforward construction of
a nonabelian world-volume action describing the dynamics
of N D$p$-branes in general background fields.
The action is composed of two parts, the Born-Infeld action
\beq
{S}_{BI}=-T_p \int d^{p+1}\sigma\,\STr\left(e^{-\phi}\sqrt{-det\left(
P\left[E_{ab}+E_{ai}(Q^{-1}-\delta)^{ij}E_{jb}\right]+
\l\,F_{ab}\right)\,det(Q^i{}_j)}
\right)\ ,
\labell{finalbi}
\eeq
with 
\beq
E_{\mu\nu}=G_{\mu\nu}+B_{\mu\nu}\ ,
\qquad{\rm and}\qquad
Q^i{}_j\equiv\delta^i{}_j+i\lambda\,[\Phi^i,\Phi^k]\,E_{kj}\ ,
\labell{extra6}
\eeq
and the Chern-Simons action
\beq
S_{CS}=\mu_p\int \STr\left(P\left[e^{i\l\,\hi_\Phi \hi_\Phi} (
\sum C^{(n)}\,e^B)\right]
e^{\l\,F}\right)\ .
\labell{finalcs}
\eeq
One of the most striking features of this result is that in the
nonabelian action, D$p$-branes couple to RR potentials
with a form degree greater than the dimension of the world-volume.
These couplings arise through inner products with the nonabelian
scalars, and vanish for the abelian theory \reef{csact}.

In both eqs.~\reef{finalbi} and \reef{finalcs}, the gauge trace
is indicated with $\STr(\cdots)$ to indicate its symmetrized nature.
The precise prescription which we propose is that inside the
trace one takes a
symmetrized average over all orderings of the $F_{ab}$, $D_a\Phi^i$,
$i[\Phi^i,\Phi^j]$, and also the individual $\Phi^i$ appearing in the
functional dependence of the background fields. This
trace prescription was adopted from matrix theory, and was required
to provide a precise agreement of the linear background couplings
for D0-branes derived from our action above, and those derived
from matrix theory \cite{wati1}. This symmetrized trace coincides with
Tseytlin's proposal for the nonabelian Born-Infeld gauge theory,
\ie eq.~\reef{finalbi} in flat space. Investigations of this proposal
have shown that it requires additional corrections involving
commutators of field strengths at sixth order \cite{notyet2,
notyet3}. However, for certain configurations, the 
flat space Born-Infeld action seems to capture the full physics
of the nonabelian Yang-Mills fields \cite{better,best}.
In any event, one should expect that
our nonabelian action will also require higher order commutator
corrections in generic situations. It would be of interest to
determine the precise limitations of our action, or rather to
determine the situations where it does reliably describe the physics.

One limitation of our analyses is that we have considered only
the bosonic fields.
There have been numerous investigations into constructing
supersymmetric world-volume actions for D-branes \cite{jhs}.
Given these results, it would, of course,
be interesting to extend the actions constructed here to incorporate
the interactions of the world-volume fermions.
In this direction, we note that the matrix theory construction
of the linear background couplings for the D0-brane
\cite{wati1} included the fermions, and so by T-duality, it
can be used to extend these couplings to arbitrary D$p$-branes
\cite{watiprep}.

Since these world-volume actions, \reef{finalbi} and \reef{finalcs},
are low energy string actions, there is an infinite series
of higher order $\alpha'$ interactions. The Born-Infeld term \reef{finalbi}
already includes an infinite set of such interactions involving
powers of $F_{ab}$, $D_a\Phi^i$ and $i[\Phi^i,\Phi^j]$. In the
abelian theory, the action \reef{biact} properly resums all the
$\alpha'$ corrections when the derivatives of the field strength
and second derivatives of the scalars vanish \cite{callan}. One still
expects further corrections involving higher derivatives of the
world-volume fields. For the nonabelian theory, the distinction
between derivatives and field strengths becomes less clear, \eg
the higher order commutator corrections \cite{notyet2}, mentioned above,
can be rewritten as higher derivative corrections. The results
of ref.~\cite{best} are an indication that the action \reef{finalbi}
does properly resum the $\alpha'$ corrections for constant self-dual
or supersymmetric field strengths.

By introducing general background fields in either the abelian
or nonabelian actions, one is implicitly also including an infinite
series of higher order $\alpha'$ corrections through the functional
dependence of these fields on the transverse directions, as
illustrated in eq.~\reef{slick}. There are still further higher
derivative corrections involving the background fields that are
independent of the transverse fields \cite{higher,anomaly,anomo}.
The focus of most investigations in this direction have been
on anomalous curvature couplings appearing in the Chern-Simons
action \cite{anomaly,anomo}. These couplings must also appear
in the nonabelian action \reef{finalcs}, but it would be interesting
to determine their precise form, in light of the commutator corrections
discovered for the nonabelian scalars.

A further comment on these anomalous couplings is that it seems
the interactions appearing in the literature \cite{anomaly,anomo}
must be incomplete. Recall that the guiding principle in constructing
eqs.~\reef{finalbi} and \reef{finalcs} was that the actions should
be consistent with T-duality. The known anomalous curvature couplings
\cite{anomaly,anomo} only consider the curvature of the metric
connection, but since T-duality exchanges components of the metric
and the Neveu-Schwarz two-form, these couplings will not be compatible
with T-duality. The consistent couplings will most likely involve
the generalized curvatures of a torsionful connection constructed
by combining the metric spin-connection and the Neveu-Schwarz
three-form field strength, \eg \cite{tort}. Recall that these
couplings arise from the anomalous transformation properties
of the world-volume spinors. So one piece of evidence in favor of
generalized curvatures appearing in these interactions is that
such a torsion term appears in the full connection of the 
super-Yang-Mills fermions of the D9-brane, \eg \cite{old}. This
makes understanding the couplings of the world-volume fermions
for general D$p$-branes an even more interesting task. 

Before leaving our discussion of the action, \reef{finalbi} and
\reef{finalcs}, we remark again that the comparison between this
action and the matrix theory action in section \ref{cheque}
made a definite prediction about that transverse five-brane charges.
That is, these elusive charges should be exactly zero, \ie
$M^{+ijklm}=0=M^{ijklmn}$. One may note that this string
theory prediction is in accord with the arguments in ref.~\cite{matbrane}
that transverse five-branes should not exist in the infinite
momentum frame of matrix theory. On the other hand, an implicit matrix theory
construction for transverse five-branes has been  suggested
using the S-duality of four-dimensional super-Yang-Mills theory
\cite{watifiv}. This is related to the observation
that while D-branes have no obvious couplings
to $\tilde{B}_{\mu\nu\l\rho\sig\tau}$, the S-duality of type IIb
superstring theory interchanges D5-branes and NS5-branes. Hence
the correct conclusion here seems to be that perturbatively $M^{+ijklm}$ and
$M^{ijklmn}$ must vanish, but that these couplings may be generated
nonperturbatively at strong coupling.

Given our nonabelian action, in section \ref{diel}, we discussed 
the D-brane analog of the dielectric effect.
When D$p$-branes are placed  in a nontrivial background
field for which the D$p$-branes would normally be regarded as neutral,
\eg nontrivial $F^{(n)}$ with $n>p+2$, new terms will be induced
in the scalar potential, and new noncommutative extrema can be
generated. At such a noncommutative extremum, the
nontrivial expectation values of the scalars will cause the
D$p$-branes to act as a multipole source for new background fields.
These ideas were explicitly illustrated by considering a system
of N D0-branes in a constant background RR field $F^{(4)}$, and
it was shown the D0-branes expand into
a noncommutative two-sphere, which carried a dipole coupling
for $F^{(4)}$. 

One should note that there are similar couplings to the
Neveu-Schwarz two-form implicit in the Born-Infeld action.
{}From the expansion of $\sqrt{det(Q)}$, one finds an interaction
of the form
\beq 
{i\over3}\lambda^2T_0\int dt\,\Tr\left(\Phi^i\Phi^j\Phi^k\right)
H_{ijk}(t)\ .
\labell{nese3}
\eeq
Hence the noncommutative ground state, in section \ref{diel}.1
for which $\Tr\left(\Phi^i\Phi^j\Phi^k\right)\ne0$, also
acts as a source of the $B$ field with
\beq
-{R_0^3\over3\pi g\ls^3}\left(1-{1\over \NN^2}\right)\int dt\,H_{123}\ .
\labell{hdipole}
\eeq
This coupling is perhaps not so surprising, as the noncommutative
ground state represents the bound state of a spherical D2-brane
and N D0-branes. Explicit supergravity solutions describing such
bound states with a planar geometry are known \cite{useful,russo},
and carry a long-range $H$ field with the same profile as the
RR field strength $F^{(4)}$. One can also derive this coupling
from the dual D2-brane formulation used in section \ref{diel}.2.
Furthermore, we observe that the presence of this coupling \reef{nese3}
means that we would find an analogous dielectric effect if
the N D0-branes were placed in a constant background $H$ field.

In section \ref{diel}.2, we compared the results of the
nonabelian D0-brane calculations to those derived from an abelian D2-brane.
The remarkable aspect of this comparison was the degree to which
the results agreed, at least in the limit of large N. This agreement
becomes even more striking when we compare the full potential
for the radial position. For the dual D2-brane, this potential
appears in the second line of eq.~\reef{dualpot}, and we rewrite
it here as
\beq
V_2(R)=\NN T_0\left(\sqrt{1+{4R^4\over\l^2 \NN^2}}-{4f\over3\l \NN}R^3\right)\ .
\labell{dualpt9}
\eeq

To calculate
the same for the D0-branes, we begin with the ansatz in
eq.~\reef{ansatz}
\beq
\Phi^i=\hat{R}\,\alpha^i_{\ssc N}
\labell{lots}
\eeq
where we will consider only the irreducible N$\times$N representation
of the SU(2) generators, but we do not fix $\hat R$ with the equations
of motion \reef{eqmot}. The corresponding physical radius 
of the corresponding noncommutative two-sphere is given by
\beq
R^2={\l^2\over \NN}\sum_{i=1}^3\Tr[(\Phi^i)^2]=\l^2\hat{R}^2(\NN^2-1)\ .
\labell{radical}
\eeq
Now the Born-Infeld potential is determined by the
determinant of the matrix $Q$ defined in eq.~\reef{extra6}, which in
the present case becomes
\beqa
det(Q_{ij})&=&det\left(\delta_{ij}
-2\l\hat{R}^2\vareps_{ijk}\al^k_{\ssc \NN}\right)
\labell{boron}\\
&=&1+4\l^2\hat{R}^4\sum_{i=1}^3(\al_{\ssc \NN}^i)^2
-8\l^3\hat{R}^6\left(\al_{\ssc \NN}^3\al_{\ssc \NN}^1\al_{\ssc \NN}^2
-\al_{\ssc \NN}^2\al_{\ssc \NN}^3\al_{\ssc \NN}^1\right)\ .
\nonumber
\eeqa
Now we make two useful observations: First, the ordering of
the matrices in the last two terms of the determinant is
ambiguous. However, in determining
the potential, this expression will appear inside the symmetrized
trace, and so by the symmetric averaging there, the contributions
of these two terms will always cancel. Hence, we can drop them
in the following calculations. The other observation is that
the SU(2) generators satisfy the relation
\beq
\sum_{i=1}^3(\al_{\ssc \NN}^i)^2=(\NN^2-1)\,I_{\ssc \NN}\ ,
\labell{casim}
\eeq
\ie this is the quadratic Casimir of the group. Hence the two terms
contributing in eq.~\reef{boron} are both proportional to the
identity, and so the gauge trace simply yields an overall factor
of N. Putting all of these results together, one finds that the
full potential arising from the nonabelian D0-brane action may
be written
\beq
V_0(R)=\NN T_0\left(1-{1\over \NN^2}\right)^{-1/2}\left(\sqrt{1-{1\over \NN^2}
+{4R^4\over\l^2 \NN^2}}-{4f\over3\l \NN}R^3\right)\ .
\labell{dualpt99}
\eeq
One may confirm that if $V_0$ is expanded for small $4R^4/\l \NN^2$,
one recovers precisely eq.~\reef{radius} as the position of the minimum.
Comparing the two potentials, \reef{dualpt9} and \reef{dualpt99}, we
now see that there will be good agreement for the two calculations
at large N using the entire potential,  not just with the
leading order expansion used in section \ref{diel}.

One way to understand this remarkable agreement is to consider the
commutator of the scalars expressed in terms of the physical radius
\reef{radical}. In terms of dimensionless
matrices, the commutator becomes
\beq
\left[\,\ls\Phi^i,\ls\Phi^j\,\right] = {i\over\pi\sqrt{\NN^2-1}}{R\over\ls}\,
\varepsilon_{ijk}\,\ls\Phi^k\ .
\labell{newcom}
\eeq
Here we see that for fixed $R$ and $\ls$, the matrices become
nearly commuting when N becomes large. Hence the nonabelian character
of the solution diminishes and so it is not surprising that the
D0-brane results match those derived from the abelian D2-brane theory.

Given the limitations of the nonabelian action discussed above,
\ie possible commutator corrections at higher orders in $F$,
one would conclude that the nonabelian
D0-brane calculations should only be trusted for small commutators.
Naively, one would expect that this would translate into restricting
the size of the noncommutative sphere to be small compared to the
string scale.  The true restriction is that
the Taylor expansion of the square root in
the potential \reef{dualpt99} converge rapidly,
and this only requires that $R<<\sqrt{\NN}\ls$. Hence for
large N, one may consider radii far larger than the string scale.

One may also argue that the radius must remain small
since the low energy action,
\reef{finalbi} and \reef{finalcs}, is simply
inapplicable if the D0-branes become  separated
by more than the string scale. That is,
if the ``low energy'' modes have masses of the
order of $1/\ls$, then one could not ignore oscillator excitations
in determining the physics of the open string sector. While the
latter is true, the radius $R$ is not the correct scale to
characterize the separation of the D0-branes. Rather one should
think of the D0-branes forming area elements of the noncommutative
two-sphere \cite{fuzz,wati2}. The typical size of the area elements
is $4\pi R^2/\NN$, and so if as above $R<<\sqrt{\NN}\ls$, the typical
separations between D0-branes is still much less than the string scale,
as required. 
The latter result can be confirmed by considering the frequency of
perturbations about the extremum \reef{radius}. One finds that 
$\omega^2=n(n+1)f^2$ for large N where $n$ is the angular
quantum number of the fluctuation mode on the noncommutative sphere.
Now if $R_0<<\sqrt{\NN}\ls$ then $f<<1/(\sqrt{\NN}\ls)$.
Hence we see that the energy of the low energy excitations is
far below the string mass scale, and so will be adequately 
described by the low energy world-volume action. 

Being able to consider large $R$ allows us to match onto the regime where
the membrane action of the dual D2-brane calculations is also applicable.
Of course, systems like the D2-brane carrying a large gauge
field flux have recently been of interest for the connections to
noncommutative gauge theories, \eg \cite{noncom1,noncom2}. Following
the analysis of ref.~\cite{noncom1}, one finds that the 
regime upon which we focussed above is precisely that in which
the standard string coupling $g$ is much smaller than the effective
string coupling $G_s$ for the theory described in terms of noncommutative
variables, \ie
\beq
g\simeq {2R^2\over\l N}G_s<<G_s\ .
\labell{noncup}
\eeq
Thus ordinary gauge theory provides the most efficient description of the
D2-branes, \ie the system is essentially abelian. It would be interesting
to try to study the system in a regime where the noncommutative
variables proved more efficient. Ultimately, one expects the D2-brane
theory to give rise to the nonabelian D0-brane physics, but making
the connection more precise should prove illuminating.

Still if the nonabelian D0- and abelian D2-brane
calculations are simply dual formulations of the same physical system,
then for large but finite N, the D2-brane framework
should still capture the small but finite noncommutative properties
found in the D0-brane calculations. Ref.~\cite{noncom1} tells us that
the difference in formulating the D2-brane theory in terms of an ordinary
or noncommutative gauge theory is in the higher derivative corrections
to the world-volume action. So perhaps the noncommutative aspects of
the bound state system will only appear in the D2-brane formalism with
the inclusion of certain higher derivative corrections to the
abelian action, \reef{biact} and \reef{csact}. Ordinarily such
terms would be ignored on the basis that their contributions give
only small corrections, but in the present case, it is precisely
small $1/\NN$ corrections that are of interest. Thus we are advocating
that resolving the discrepancies between the dual pictures
requires determining corrections to {\it both} formalisms, \eg
higher order commutator corrections for the D0-branes, and
higher derivative corrections for the D2-branes.

Finally, after arguing the merits of our calculations in section \ref{diel},
we must still acknowledge one significant shortcoming --- namely the
supergravity background is not a consistent background solution
of the type IIa supergravity equations of motion. One could argue that
this background might be approximated by placing the D0-branes
in the asymptotic region far from a D2-brane, and focusing
on a small region in which the gradients of the background fields
can be neglected. One can not take seriously all of the details
of the calculations. For example, the potentials \reef{dualpt9}
and \reef{dualpt99} are in fact unstable since they are dominated
by the $R^3$ term at large radius. However, this instability
will only become apparent at radii of the order of $1/f$, but the
gravitational effects of the total energy of the four-form
enclosed in such a region will be of order 1. Hence the gravitational
back-reaction cannot be neglected on this scale, and so we
conclude that this instability is simply an artifact of not
chosing a consistent solution for the background.

However, the calculations in section \ref{diel} were presented
in the spirit of a toy calculation to illustrate the effects
of the new nonabelian interactions appearing in the
action, \reef{finalbi} and \reef{finalcs}, and in particular
to demonstrate the dielectric effect for D-branes. 
A more detailed examination of these issues 
for realistic backgrounds is in preparation \cite{prep}.

\section*{Acknowledgments}
This research was supported by NSERC of Canada and Fonds FCAR du Qu\'ebec.
I would like to acknowledge useful conversations with Neil Constable, Clifford
Johnson, Aki Hashimoto, Joe Polchinski,
Oyvind Tafjord, Washington Taylor and Mark van
Raamsdonk. Research at the ITP, UCSB was supported by NSF Grant PHY94-07194.

\noindent While this paper was in the final stages of preparation, I was
informed by Washington Taylor and Mark van Raamsdonk of their recent work
\cite{watiprep}, which has significant overlap with the material
in sections 3 and 4.

\end{document}